%% file: main.tex
\documentclass[preprint,12pt,3p,authoryear]{elsarticle}  

\usepackage{graphicx}
\usepackage{appendix}
\usepackage{txfonts}
\usepackage{siunitx}
\DeclareSIUnit\parsec{pc}
\DeclareSIUnit\year{yr}
\DeclareSIUnit\magnitude{mag}
\usepackage[hidelinks]{hyperref}
\usepackage{xurl}
\usepackage{booktabs}
\usepackage{xcolor}
\usepackage{soul}
\usepackage{nicefrac}

\ifdefined\qty\else
  \ifdefined\NewCommandCopy
    \NewCommandCopy\qty\SI
  \else
    \NewDocumentCommand\qty{O{}mm}{\SI[#1]{#2}{#3}}
  \fi
\fi

\journal{Astroparticle Physics}

\begin{document}

\begin{frontmatter}
\title{The influence of the 3D Galactic gas structure on cosmic-ray transport and $\gamma$-ray emission}

%
  % main contributors in order of decreasing contribution
  \author[1]{ Andrés~Ramírez}
  \ead{Andres.Ramirez-Tapias@uibk.ac.at}
  % secondary contributors alphabetically
  \author[2,3]{Gordian~Edenhofer}
  \author[2,3,6]{Torsten~A.~Enßlin}
  \author[2]{Philipp~Frank}
  \author[4]{Philipp~Mertsch}
  \author[5]{Vo~Hong~Minh~Phan}
  \author[4]{Laurin~Söding}
  \author[2,3]{Hanieh~Zandinejad}
  % supervisor of major contributor at the end
  \author[1]{Ralf~Kissmann}

\affiliation[1]{
  organization={Universität Innsbruck, Institut für Astro- und Teilchenphysik},
  addressline={Technikerstr. 25/8}, 
  city={Innsbruck},
  postcode={6020},
  country={Austria}}
  \affiliation[2]{
  organization={Max Planck Institute for Astrophysics},
  addressline={Karl-Schwarzschild-Straße 1}, 
  city={Garching bei München},
  postcode={85748},
  country={Germany}}
\affiliation[3]{
  organization={Ludwig Maximilian University of Munich,},
  addressline={Geschwister-Scholl-Platz 1}, 
  city={München},
  postcode={80539},
  country={Germany}}
\affiliation[4]{
  organization={Institute for Theoretical Particle Physics and Cosmology},
  addressline={RWTH Aachen University, Sommerfeldstr. 16}, 
  city={Aachen},
  postcode={52074},
  country={Germany}}
\affiliation[5]{
  organization={Sorbonne Université, Observatoire de Paris, PSL Research University},
  addressline={LERMA, CNRS UMR 8112}, 
  city={Paris},
  postcode={75005},
  country={France}}
\affiliation[6]{
  organization={Excellence cluster ORIGINS},
  addressline={Boltzmannstr. 2}, 
  city={Garching},
  postcode={85748 },
  country={Germany}}

\begin{abstract}
  Cosmic rays (CRs) play a major role in the dynamics of the interstellar medium (ISM). Their interactions and transport ionize, heat, and push the ISM thereby coupling different regions of it. The spatial distribution of CRs depends on the distribution of their sources as well as the ISM constituents they interact with, such as gas, starlight, and magnetic fields. Particularly, gas influences CR fluxes and $\gamma$-ray emission. We illustrate the influence of realistic and largely structured 3D gas distributions on CR transport and $\gamma$-ray emission, by studying their correlation using the PICARD code and multiple samples of recent 3D reconstructions of the HI and H2 Galactic gas constituents. We adjust the diffusion coefficient $D_{xx}$ and Alfvén speed $v_{\text{A}}$ to reproduce local measurements of B/C abundances and find that these parameters depend non-linearly on the local distribution of gas. When simulating CR transport, the distributions of CR fluxes exhibit energy-dependent structures that vary for all CR species due to their corresponding loss processes. Regions of enhanced secondary (primary) species are spatially correlated (anti-correlated) with the gas density. Furthermore, we show that the morphology of gas clouds alone impacts CR flux predictions. For $\gamma$-ray emission, we observe a high sensitivity of the $\gamma$-ray emissivities to gas structures, as these determine the spatial distributions of hadronic interactions and bremsstrahlung. This way, we have for the first time calculated how well-defined uncertainties in a structured gas model propagate to CR transport and $\gamma$-ray emission.
\end{abstract}
% conclusions heading (optional)
{}

\end{frontmatter}
\input{Sections/Introduction}
\input{Sections/Methodology}
\input{Sections/Results}
\input{Sections/Conclusion}
\section*{Acknowledgements}
  Data used in this study was extracted with the help of the CRDB cosmic-ray database \citep{Maurin2014,Maurin2020db,Maurin2023db}.
  All figures in this publication have been created using \emph{matplotlib}~\citep{Hunter2007}, where the healpy and HEALPix package were additionally used for some.
  This research was funded in part, by the Austrian Science Fund (FWF) 10.55776/I5925. For open access purposes, the author has applied a CC BY public copyright license to any author accepted manuscript version arising from this submission.
  Part of this work was supported by the German \emph{Deut\-sche For\-schungs\-ge\-mein\-schaft}, DFG project number 495252601.
  Gordian Edenhofer acknowledges that support for this work was provided by the German Academic Scholarship Foundation in the form of a PhD scholarship (``Promotionsstipendium der Studienstiftung des Deutschen Volkes''). Philipp Frank acknowledges funding through the German Federal Ministry of Education and Research for the project ErUM-IFT: Informationsfeldtheorie f\"ur Experimente an Gro\ss forschungsanlagen (F\"orderkennzeichen: 05D23EO1). Vo Hong Minh Phan acknowledges support from the Initiative Physique des Infinis (IPI), a research training program of the Idex SUPER at Sorbonne Universit\'e. The authors would like to thank Stefan Lepperdinger for his contribution in the early stages of this work.
\bibliographystyle{elsarticle-num-names}
\bibliography{references}
\appendix
\input{Sections/appendixA}
\input{Sections/appendixB}
\input{Sections/appendixC}

\end{document}

%% file: Sections/Introduction.tex
\section{Introduction}
\label{introduction}

Cosmic rays (CRs) are an important component of the Galactic interstellar medium (ISM). They interact with other constituents of the ISM such as gas, dust, turbulent magnetic fields and interstellar radiation. Furthermore, CRs are believed to affect the chemical and dynamical evolution of the ISM from small (star-forming) to large, Galactic scales \citep{padovani2020,gabici2022,phan2023,simpson2023}. Several studies suggest that CRs drive Galactic winds and, thus, could play an essential role in the evolution history of the Milky Way (MW) \citep{farcy2022,girichidis2022}.

Despite the importance of CRs in the Galactic ecosystem, the transport of these particles is still far from being fully understood. Analyses of CR secondary-to-primary ratios (e.g. B/C and $^{10}$Be/$^{9}$Be) indicate that these particles propagate diffusively within a magnetized halo (with a typical height of around 3-6\,kpc) surrounding the Galactic disk \citep{evoli2020,maurin2022,jacobs2023}. The diffusion of CRs is thought to be due to their interactions with magnetic turbulence (the turbulent component of the Galactic magnetic field). In this respect, Galactic CR modelling formally follows from coupling the transport of CRs to magnetic turbulence \citep{Schlickeiser2001,blasi2013}, using a set of non-linear coupled differential equations. Solving these coupled equations remains challenging in many different ways. For example, it is not yet clear whether magnetic turbulence is predominantly generated from the cascade of large-scale turbulence injected by supernova explosions or from CRs themselves via the streaming instability \citep{skilling1975,nava2016,jacobs2022}. Several recent analyses, in fact, indicate that both mechanisms are important \citep{thomas2019,thomas2020} and in different energy ranges they lead to features in the CR spectra \citep{genolini2017,evoli2018}. Another difficulty in modelling Galactic CR transport and magnetic turbulence self-consistently is our lack of knowledge on the large-scale magnetic field of the MW \citep{beck2015}. More precise 3D models of the Galactic magnetic field \citep{Jansson20121,Jansson20122} would be essential for a better understanding of CR propagation \citep{haverkorn2019} since the diffusion coefficients of CRs parallel and perpendicular to the field lines are expected to differ, making the Galactic distribution of CRs sensitive to the exact geometry of the large-scale magnetic field \citep{cerri2017,mertsch2020,giacinti2023}. 

Given all the above-mentioned complexities, the propagation of CRs is normally modelled in a linear fashion, via the CR transport equation assuming parametric forms of the diffusion coefficients and other transport and acceleration parameters such as the advection speed (due to plasma waves or Galactic winds) or the diffusion coefficient in momentum space (for the process of re-acceleration). In fact, the growing amount of high-precision CR data has led to the development and refinement of multiple sophisticated CR transport and interaction codes such as GALPROP \citep{Porter_2022}, DRAGON \citep{Evoli_2017}, USINE \citep{MAURIN2020106942} and PICARD \citep{KISSMANN201437}. These codes provide numerical or semi-analytical solutions for the CR transport equation, taking into account many different physical processes not only for transport but also for particle acceleration and interactions between CRs and the ISM (resulting in energy loss and production of secondary CRs via spallation). The combination of these transport codes and high-precision CR data \citep{PAMELAData,CALETData,AMS02Data}, covering energy ranges from the MeV to the TeV scale and beyond, has been extensively used to extract corresponding transport and acceleration parameters \citep{Strongannurevnucl57090506123011,grenier2015,gabici2019}. 

Many models using such codes use axisymmetric Galacto-centric gas distributions to simulate CR transport \citep[see e.g.,][]{Evoli_2012,Gaggero_2015,Orlando2019,Widmark2023}. Since gas serves as a target for spallation reactions of primary CRs producing secondary CRs, approximate 2D gas distributions and more realistic 3D ones are expected to yield different estimates for secondary-to-primary ratios for otherwise identical transport parameters. Conversely, different estimates of the transport parameters are expected between 2D and 3D gas models when inferring them from data on CR secondary-to-primary ratios. This is observed when a more structured 3D model is used for the inference of transport parameters instead of a smooth axisymmetric model, yielding a lower diffusion coefficient $D_{xx}$ and a reduced Alfvén speed $v_{\text{A}}$ (see e.g.,~\citet{Johannesson_2018}, but also Table \ref{tab:transportparameters} here).

Furthermore, interactions between CRs and gas can result in observable diffuse $\gamma$-ray emissions, i.e.\ via decays of $\pi^0$ (produced in hadronic interactions of CRs with interstellar gas nuclei) or via bremsstrahlung by CR electrons. Diffuse $\gamma$-ray data can, therefore, be used to derive the Galactic distribution of CRs assuming models for the 3D gas distributions \citep{ackermann2012,tibaldo2021}. Despite their relevance, 3D gas distributions are still not commonly used in studies of Galactic CR transport. 

One of the main methods for deriving 3D gas distributions is to analyze data of spectral emission and absorption lines. Atomic hydrogen (HI), for instance, can be traced through the 21-cm emission line. Since gas at different locations along a line of sight in general has different relative velocities with respect to the observer due to Galactic rotation, it emits this line at different Doppler-shifted frequencies. This means that, given a model for Galactic rotation, data of this emission line can be used to reconstruct a model for the 3D distribution of HI. A similar approach can be applied to the $J=1 \rightarrow 0$ emission line of $^{12}$CO to trace molecular hydrogen (H$_2$) assuming a conversion factor between $^{12}$CO and H$_2$ gas density. However, given our vantage point in the Galaxy, many of these reconstructed gas maps suffer from a number of ambiguities. For instance, the peculiar motion of gas (i.e.\ random motions of gas, on top of the large-scale rotational gas flow caused e.g. by stellar winds and supernova explosions) introduce artificial structures stretching along the lines of sight into the 3D gas reconstructions, known as the finger-of-god effect \citep[see e.g.,][]{nakanishi2003,nakanishi2006,pohl2008}. Recently it has been suggested that this problem can be cured by taking the spatial correlations of gas into account \citep{philipp_mertsch_2020_5501196,philipp_mertsch_2022_5956696}. Incorporating spatial correlations into 3D gas reconstructions turns those into high-dimensional Bayesian inference problems. Such have been addressed via information field theory \citep{ift2009,ift2019} in \citet{philipp_mertsch_2020_5501196} and \citet{philipp_mertsch_2022_5956696} by using the \texttt{NIFTy} package\footnote{\href{https://github.com/NIFTy-PPL/NIFTy}{\texttt{github.com/NIFTy-PPL/NIFTy}}}\citep{nifty1,nifty3,nifty5,niftyre}. This has not only provided improved 3D gas reconstructions but also quantified their remaining uncertainties. An important feature of these gas reconstructions, when compared to analytical 3D models, is their higher dynamic range (i.e. local densities vary in a wider range) which leads to a higher contrast between over- and under-density regions.

In this work, we study how 3D gas structures, as seen in such improved 3D gas reconstructions, influence CR transport and $\gamma$-ray emission. We model this by introducing the 3D reconstructed maps and an axisymmetric averaged version of the same 3D maps into CR transport simulations.
We use both models to correlate gas densities to transport parameters (i.e. Diffusion coefficient and Alfvén speed), and to simulate the energy-dependent spatial distribution of CRs and of $\gamma$-ray emission with the PICARD code. In contrast to previous works \citep[e.g.][]{Johannesson_2018}, rather than aiming for a best-fit model for each investigated setup, we focus on a model-to-model comparison to quantify the impact of localized gas structures on the simulated distributions of CR fluxes and $\gamma$-ray emissivities. In particular for the latter, we ensure the same 3D gas distribution is used to perform the line-of-sight integration leading to all-sky $\gamma$-ray emission maps.

The structure of the remainder of this work is as follows. In section \ref{sec:Setup}, we detail how the reconstruction and axisymmetric models are introduced into the PICARD simulation setups. Then, in section \ref{sec:Results and Discussion} we discuss the morphological features induced by the 3D gas on simulated CR transport and $\gamma$-ray emission, and compare the resulting CR spectra and transport parameter estimates between 2D ring and 3D models. Finally, in section \ref{Summary and conclusions} we summarize our results and present our conclusions.

%% file: Sections/Methodology.tex
\section{Methodology}
\label{sec:Setup}

\begin{figure*}
	\centering
	\includegraphics[width=\textwidth]{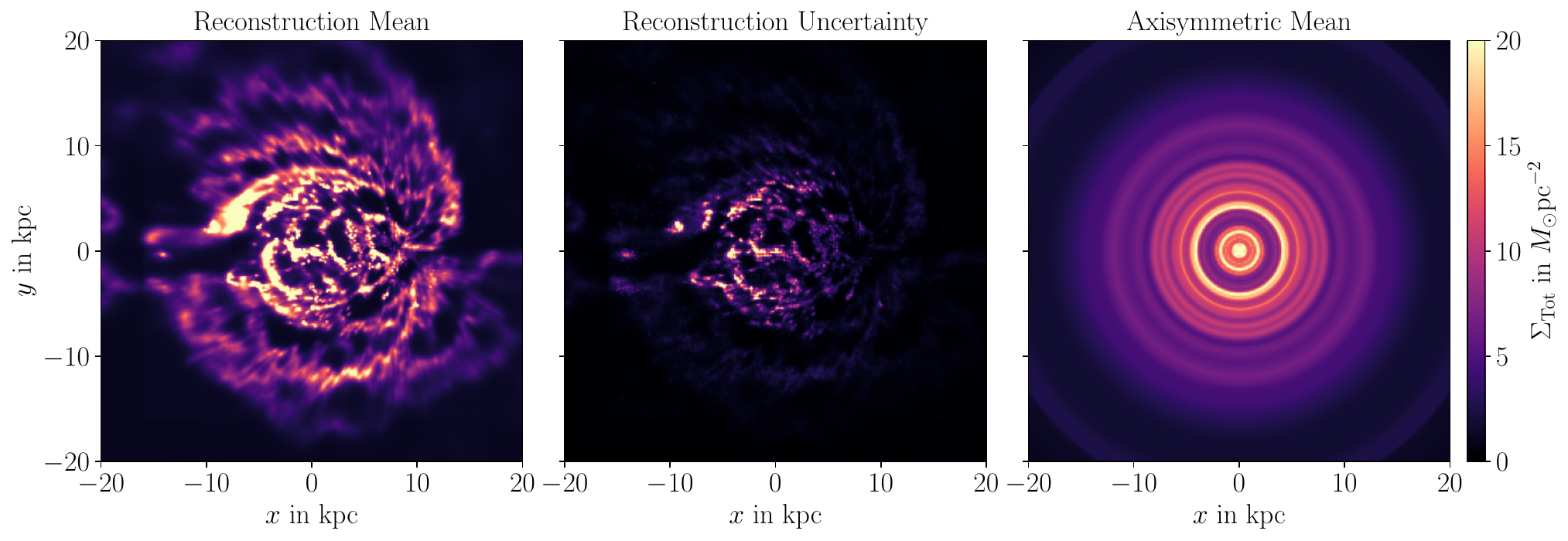}
	\caption{\textit{Left:} surface mass density for the reconstruction mean map of the 3D gas model. \textit{Middle:} corresponding uncertainty for the constituent sample maps. \textit{Right:} surface column density for the axisymmetric map of the 2D model.}
	\label{totprot}
\end{figure*}

\begin{table*}
	\caption{Transport parameters used for the two different setups.}
	\label{tab:transportparameters}
	\begin{center}
	\begin{tabular}{lcc} \toprule 
		Parameter & 2D model & 3D model \\
		\midrule 
		\textit{ General}& \\
		Halo height [kpc] & \multicolumn{2}{c}{4} \\
		Galactic radius [kpc] &  \multicolumn{2}{c}{20} \\
		Diffusion coefficient $D{xx}~^1$ [$10^{24}$m$^2$/s] & 5.10 & 3.90 \\
        Reference Rigidity $R_0$ [GV] & \multicolumn{2}{c}{4} \\
		Diffusion coefficient index $a_1$ below $R_0$ & $0.26$ & $0.21$ \\
        Diffusion coefficient index $a_2$ above $R_0$ & $0.38$ & $0.40$ \\
		Alfvén speed $v_{\text{A}}$ [$10^3$\,m/s] & $31.0$ & $21.0$ \\
		\textit{ Nuclei injection spectrum} & \\
		Index below  break & \multicolumn{2}{c}{1.89}\\
		Index above break & \multicolumn{2}{c}{2.39}\\
		Break energy [GeV] & \multicolumn{2}{c}{11.41} \\
		\textit{ Nuclei normalisation} &\\
		Normalisation energy [GeV] & \multicolumn{2}{c}{108} \\
		Normalisation flux$~^\mathrm{a}$ &  \multicolumn{2}{c}{$4\cdot10^{-2}$}  \\
		\textit{ Electron injection spectrum} & \\
		Index below $1^{\mathrm{st}}$ break &  \multicolumn{2}{c}{$1.6$} \\
		Index between $1^{\mathrm{st}}$ and $2^{\mathrm{nd}}$ break & \multicolumn{2}{c}{$2.425$} \\
		Index above $2^{\mathrm{nd}}$ break & \multicolumn{2}{c}{4.0} \\
		$1^{\mathrm{st}}$  break energy [GeV] & \multicolumn{2}{c}{1} \\
		$2^{\mathrm{nd}}$ break energy [GeV] & \multicolumn{2}{c}{$2.5\cdot10^3$} \\
		\textit{ Electron normalization} &\\
		Normalisation energy [GeV] & \multicolumn{2}{c}{25}\\
		Normalisation flux$~^\mathrm{a}$ & \multicolumn{2}{c}{$1.09\cdot10^{-2}$}\\
		\bottomrule		
        \multicolumn{3}{l}{$^1~~D{xx}=\beta (R/R_0)^a$} \\
        \multicolumn{3}{l}{$^\mathrm{a}$\, Normalisation flux given in units of [m$^{-2}$s$^{-1}$sr$^{-1}$(GeV/nucleon)$^{-1}$]} \\
		\end{tabular}
		\end{center}
\end{table*}

In the PICARD code, a steady state solution to the CR transport equation for each nucleus is obtained on a numerical grid by either using an axisymmetric cylindrical coordinates or a three-dimensional Cartesian grid in space -- as described for our models below -- and one energy dimension, where we use a logarithmically equidistant axis, ranging from 10\,MeV to 1\,PeV in 127 grid points.

Our setup is based on a previous axisymmetric model from \citet{ackermann2012}, i.e. using the short-hand notation introduced therein, the model $^{\text{S}}$Y$^{\text{Z}}$4$^{\text{R}}$20$^{\text{T}}$150$^{\text{C}}$5. This designator refers to a halo size of $z_h=4$\,kpc, radial boundaries at $R_h=20$\,kpc, spin temperature $T_S=150$\,K, and $E(B-V)$ magnitude cut of $c=5$.
In particular, we use the axisymmetric cosmic-ray source distribution by \citet{YusifovKuecuek2004AnA422_545}, which we truncate beyond a Galactocentric radius of 15\,kpc as also discussed in \citet{KISSMANN201539}.
The transport parameters are the same as in $^{\text{S}}$Y$^{\text{Z}}$4$^{\text{R}}$20$^{\text{T}}$150$^{\text{C}}$5 apart from the Alfvén speed, the energy dependence of spatial diffusion, and the particle normalizations, as discussed below.
Additionally, the models for the magnetic field and the radiation field are also axisymmetric as also used in \citet{ackermann2012}.
Correspondingly, the new three-dimensional gas reconstructions are the only non-axisymmetric components considered in the transport equation.
We use the same set of spallation cross-sections and energy loss process as GALPROP v54\footnote{\href{https://galprop.stanford.edu/download/manuals/galprop_v54.pdf}{\texttt{galprop.stanford.edu/download/manuals/galprop\_v54.pdf}}} \citep[see e.g.,][]{StrongMoskalenko1998ApJ509_212, MoskalenkoStrong1998ApJ493_694, MoskalenkoEtAl2002ApJ565_280}.

Nuclei are coupled through spallation reactions which lead to corresponding loss and source terms. For the handling of the related nuclear network see \citet{KISSMANN201539}. Thus, for particles resulting from interactions with Galactic gas, $n_{\rm Tot}$ defines the interaction regions producing secondaries.

We adapt the recently reconstructed 3D maps of molecular (H$_2$) -- based on the reconstruction of Galactic CO with a fixed $X_{CO}=2\times10^{20}$\,cm$^{-2}$(K\,km\,s$^{-1}$)$^{-1}$ -- \citep{philipp_mertsch_2020_5501196} and atomic (HI) \citep{philipp_mertsch_2022_5956696} hydrogen into the PICARD simulations. These reconstructions were performed following two different gas flow models, namely BEG03 (based on the results of a smoothed particle hydrodynamics simulation) and SBM15 (based on a semi-analytical model for gas-carrying orbits) as introduced in the original articles. Since we are interested in the distribution of 3D structures, both are equally good candidates for the purpose of this work. Here, we selected the HI and H$_2$ maps based on the BEG03 gas flow model. These maps come in the form of 32 H$_2$ and 16 HI independent posterior sample maps. Out of the 512 possible combinations of H$_2$ and HI sample pairs, we subsample by randomly choosing 20 map pairs to account for fluctuations in the gas structure. %(see appendix \ref{appendix:samples} for details about the significance of the chosen subsample). 
To load each pair of gas maps into the simulations, we interpolate them from their original reconstruction grid onto the PICARD simulation grid. The gas maps are then added with the analytical model of ionised (HII) gas used in PICARD to generate the total distribution of protons in the galaxy $n_{\rm Tot}$. This is calculated as

\begin{equation}
    n_{\rm Tot}(\vec{r})=2n_{\text{H}_2}(\vec{r})+n_{\text{HI}}(\vec{r})+n_{\text{HII}}(\vec{r}).
\end{equation}

We assume that the distribution of ISM He follows the one of ISM H, i.e. we compute the corresponding distribution assuming the number density of He to be a fraction of 11\% of $n_{\rm Tot}$. From these maps we produce two models. For the first one, from hereon referred to as the 2D model, we average over the selected sample maps to produce a reconstruction mean map. We interpolate this map from the reconstruction grid to a cylindrical grid $(r,\phi,z)$ with the same resolution in $r$ as that of $x$ and $y$ in the original reconstruction grid, as well as the same resolution in $z$, and 720 grid points defining the $\phi$ axis. To produce an axisymmetric 2D model, we average the result over the azimuth $\phi$ and store the averaged value at the corresponding $(r,\phi,z)$ grid point for every $\phi$. The resulting distribution is then interpolated to a 3D Cartesian simulation grid. For the second setup, from hereon referred to as the 3D model, we interpolate the 20 samples directly onto the same Cartesian simulation grid $(x,y,z)$. The Cartesian simulation grid spans from -20\,kpc to 20\,kpc with 257 grid points in both $x$ and $y$ dimensions, and from -4\,kpc to 4\,kpc with 65 grid points in $z$. 

 Given the multiple samples we study here, higher resolution setups become prohibitively expensive. To test the impact of the chosen spatial resolution, we investigate the differences between predictions using our default grid and a finer one, using twice the spatial-resolution (i.e. 513 grid points in $(x,y)$ and 129 grid points in $z$). We compare the simulated distributions of primary and secondary protons using both grids at the two main locations of interest for this work: Earth and the Galactic center. On the coarse grid, we find that the fluxes are on average higher by ~2.5\% without any changes to the spectral slope. Deviations are larger only at very localized gas clouds, raising to ~10\%, due to the shape of these clouds blurring out in the lower resolution. However, structures are not principally different. Thus the coarse grid proves sufficient for the purpose of this work.

Both models are illustrated in Fig \ref{totprot} where we show the resulting surface mass density ($\Sigma_\mathrm{Tot}$) of the reconstruction mean map and the uncertainty (e.g one standard deviation) computed over the samples, as well as the surface mass density of the 2D model map.

Different gas models cause the simulated secondary to primary ratios at Earth to differ when compared to data (i.e. B/C ratio data). To understand these differences, we heuristically calibrate our transport parameters by modifying those related to the strength and speed of CR diffusion. Namely, the diffusion coefficient $D_{xx}$, its corresponding indices $a_1$ and $a_2$, and the Alfvén speed $v_{\text{A}}$.  We refrain from fitting many transport parameters for each model, because we are mainly interested on the impact of the gas only. We perform this calibration by running multiple simulations with slightly different values for the parameters thereof, until finding a rough by-eye agreement between the simulated B/C ratios and AMS-02 data \citep{AMS02Data2}. We do this for both the 2D and 3D models using the axisymmetric and reconstruction mean maps shown in Fig. \ref{totprot}, respectively. In this work, the heaviest element considered in the simulations is Silicon, where contributions from heavier elements are neglected, as they impact the distribution of Carbon and Boron isotopes by less than a percent. The resulting set of transport parameters are presented in Table \ref{tab:transportparameters}. 

Diffuse $\gamma$-ray emission is computed considering the contributions from $\pi^0$-decay, bremsstrahlung, and IC scattering. For this calculation, transport results for relevant CR species (hydrogen, helium, electrons and positrons) are used with the corresponding cross-sections to generate emissivities at the specified energies of interest \citep[see e.g,][]{StrongMoskalenko1998ApJ509_212,StrongEtAl2000ApJ537_763}. For $\pi^0$-decay and bremsstrahlung, specific emissivities,  using ISM H and ISM He as target, are produced which are later multiplied with the total proton distribution to obtain the final $\gamma$-ray emissivities in the Galaxy. For these simulations, the spatial grid is the same as in the transport step and CR energies range from 10\,GeV to 1\,TeV in 10 logarithmically equidistant grid points. After the diffuse emission is computed, the specific emissivities for $\pi^0$-decay and bremsstrahlung are interpolated onto a Cartesian grid with the same spatial resolution as the sample maps, to multiply them with the total gas distribution. Such grid preserves the full resolution of the sample maps, spanning the same dimensions as that of the 3D model, with (640,640,128) grid points in the ($x,y,z$) dimensions. From the local emission, all-sky $\gamma$-ray emission follows via line-of-sight (LOS) integration.

%% file: Sections/Results.tex
\section{Results and Discussion}
\label{sec:Results and Discussion}

The main results of our simulations are the energy-dependent spatial distributions of Galactic CR fluxes ($J(E)$) and of local $\gamma$-ray emissivities ($\epsilon(E)$) -- as well as the LOS integrals of the latter -- computed with the 3D model. In this section, we present the summary statistics of these results and compare them, where applicable, to the morphology of the gas reconstruction maps and the results obtained by using the 2D model. All uncertainties shown here correspond to the 1-$\sigma$ region of the simulation outputs.
For all our results related to CR nuclei fluxes we define the energy axis by their corresponding kinetic energy per nucleon (n). We denote this convention as $E_{\rm kin}$ in GeV/n units. For leptons, we use their corresponding kinectic energy $E_{\rm kin}$ in GeV units.

\subsection{Impact on Transport Parameters}
\label{subsec:B/C Ratio Fit}

\begin{figure}
	\centering
	\includegraphics[scale=0.7]{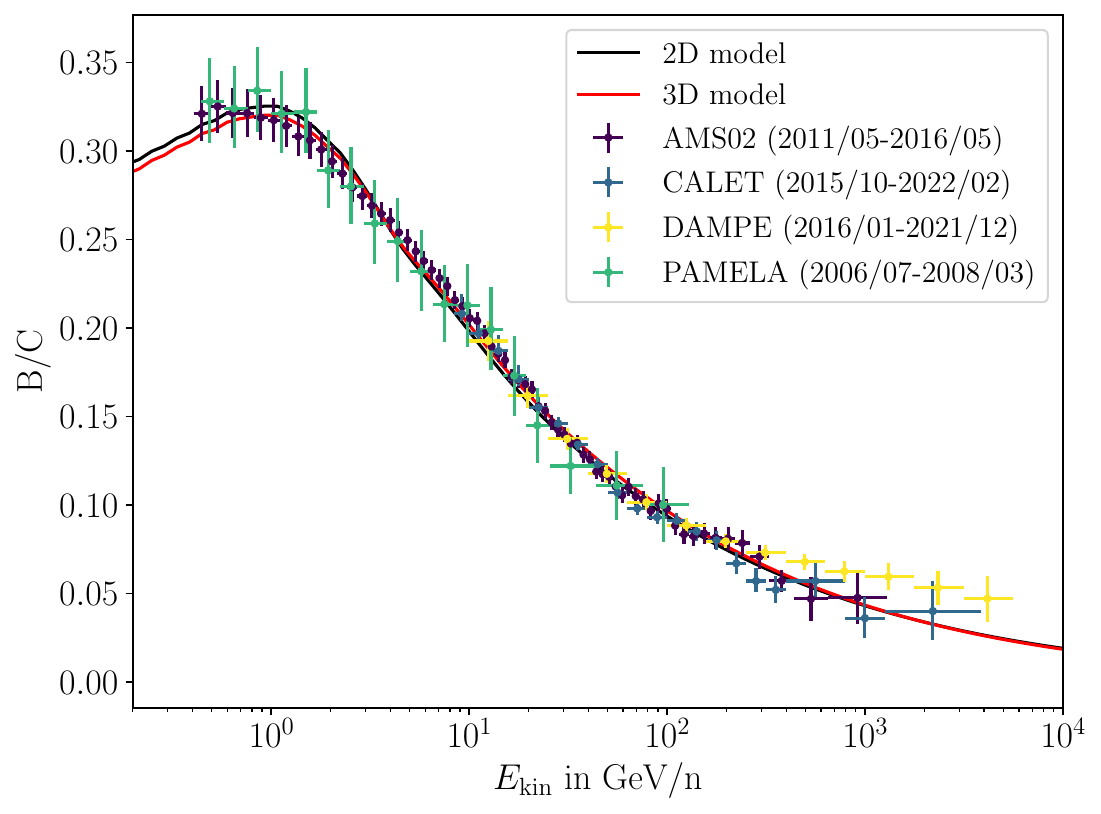}
	\caption{B/C ratio fit for 2D and 3D models at Earth compared to AMS-02 \citep{AMS02Data2}, CALET \citep{CALETData}, DAMPE \citep{DAMPEData} and PAMELA \citep{PAMELAData} data. The ratio is shown with a heliospheric modulation of 678.6 MV estimated for the AMS-02 measurement.}
	\label{BCfit}
\end{figure}

\begin{figure*}
	\centering
	\includegraphics[width=\textwidth]{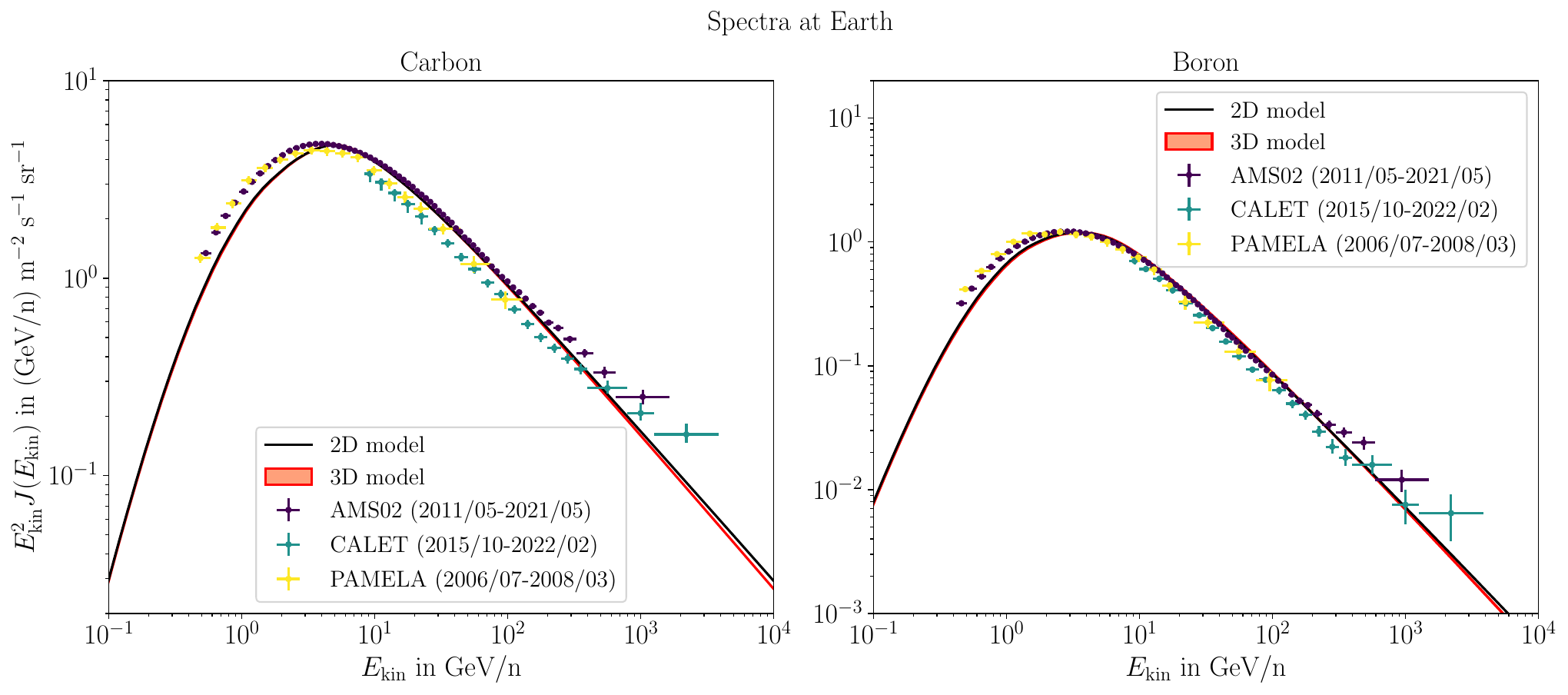}
	\caption{Carbon and Boron spectra for the 2D and 3D models at Earth compared to AMS-02 \citep{AMS02Data}, CALET \citep{CALETData} and PAMELA \citep{PAMELAData} data. The spectra are shown with a heliospheric modulation of 587.9\,MV estimated for the AMS-02 measurement.}
	\label{TotBCfit}
\end{figure*}

Boron nuclei are predominantly produced via the interaction of heavier nuclei (e.g. Carbon and oxygen) with ISM gas. The relative abundances between secondary and primary nuclei are sensitive to the mechanisms of CR transport, making their measurement an important probe to constrain the parameters involved in CR propagation \citep{strong2007CRP,DAMPECOLLABORATION20222162}. Particularly, carbon and boron abundances are extensively well documented. Therefore, we use the available B/C ratio data to investigate the constraints introduced in the transport parameters related to diffusion, given the usage of the gas reconstructions. In Fig \ref{BCfit} we show the fits to the B/C ratio data resulting from the calibration of the transport parameters described in section \ref{sec:Setup}. We quantify our goodness of fit for energies above 10 GeV to avoid the impact of heliospheric modulation, which we only consider via the force-field approximation. We find a $\chi^2$ value of 1.47. We note that the 2D model requires $D_{xx}$ and $v_{\text{A}}$ values slightly  below those used for the $^{\text{S}}$Y$^{\text{Z}}$4$^{\text{R}}$20$^{\text{T}}$150$^{\text{C}}$5 model in \citet{ackermann2012} where $D_{xx}=5.28\times 10^{24}$\,m$^2$\,s$^{-1}$ and $v_{\text{A}}=33.1$\,km\,s$^{-1}$. 

The transport parameters of the 3D model deviate significantly from the ones in the 2D model by requiring a smaller diffusive strength to remain close to the AMS-02 data. This behaviour is related to the gas density near Earth in both models. We check this behavior in each model by evaluating the corresponding average gas densities $\bar{n}$ within a sphere of 4kpc radius around the location of Earth. For the 2D model, $\bar{n}_{\rm{2D}}(|\vec{r}-\vec{r}_{\rm{Earth}}|\leq 4 \rm{kpc})\approx62\times10^{-3}$\,cm$^{-3}$ and for the 3D model, $\bar{n}_{\rm{3D}}(|\vec{r}-\vec{r}_{\rm{Earth}}|\leq 4 \rm{kpc})\approx53\times10^{-3}$\,cm$^{-3}$. We find that $\bar{n}_{\rm{2D}}\approx1.17\bar{n}_{\rm{3D}}$ in this region, leading to ${D_{xx}}_{\rm{2D}}\approx 1.31{D_{xx}}_{\rm{3D}}$ and ${v_{\text{A}}}_{\rm{2D}} \approx 1.48 {v_{\text{A}}}_{\rm{3D}}$. Consequently, we observe a direct correlation between the average gas density in the vicinity of Earth and with $D_{xx}$ and $v_{\text{A}}$, where smaller average gas densities require weaker diffusion and reacceleration. 
However, we note that changes in $n_{\rm{3D}}$ do not translate linearly to changes in $D_{xx}$ and $v_{\text{A}}$, this results from gas structures at different scales affecting the flux of CRs via their interactions (e.g. spallation, energy losses), which we discuss further in Section \ref{subsec:correlations}. We also note that in both models the diffusion coefficient indices $a_1$ and $a_2$ differ (i.e. $|a_1-a_2|_{2D}\approx0.63|a_1-a_2|_{3D}$) and deviate from the standard dependence for Kolmogorov turbulence where $a=0.33$. Similarly to $D_{xx}$ and $v_{\text{A}}$, the energy dependence of the cross sections used to parameterize the CR spallation and energy losses, when folded with either gas model, leads to a change in the slope of the simulated CR spectra. We address such a slope change by adjusting $a_1$ and $a_2$ to fit our simulated B / C ratio to the AMS-02 data. As discussed above, $\bar{n}_{\rm{2D}}>\bar{n}_{\rm{3D}}$ in the 4kpc spherical region centered on Earth, since $|a_1-a_2|_{2D}<|a_1-a_2|_{3D}$, we note that the quantity $|a_1-a_2|$ is anti-correlated to $\bar{n}$. Furthermore, the distribution of local gas clouds -- even for the same local $\bar{n}$ among gas realizations -- also affects the simulated CR flux, indirectly impacting the choice of transport parameters. We discuss this relation in more detail in section \ref{subsec:Cosmic-Ray Distributions and Spectra}.

In Fig \ref{TotBCfit}, we present the simulated carbon and boron spectra at Earth that constitute the fitted B/C ratio, as well as corresponding observations. The main deviations between data and simulated spectra are present for energies below 10\,GeV/n, where at the estimated heliospheric modulation, the simulated fluxes are below the data. Stable Boron isotopes (i.e. ${}^{10}$B and ${}^{11}$B) are mainly produced from spallations of CR ${}^{12}$C and of CR ${}^{16}$O with ISM protons, which also produce unstable Carbon isotopes that decay further (see Table 1 in \citet{HeinbachSimon99} for the related cross-sections). This anti-correlates the population of carbon and boron non-linearly, where the latter is dominated by the ${}^{12}$C (see e.g. Figs \ref{BCSpectra1} and \ref{BCSpectra2}). As a result, given enough local gas density, we expect a greater gain of Boron nuclei from a lower loss of Carbon nuclei. Therefore, the relative deviations from the data being larger for Carbon than for Boron result from the sensitivity of the distribution of CRs to both the choice of transport parameters and the local gas density. 

\subsection{Impact on CR distributions}
\label{subsec:Cosmic-Ray Distributions and Spectra}

\begin{figure*}
	\centering
	\includegraphics[scale=0.38]{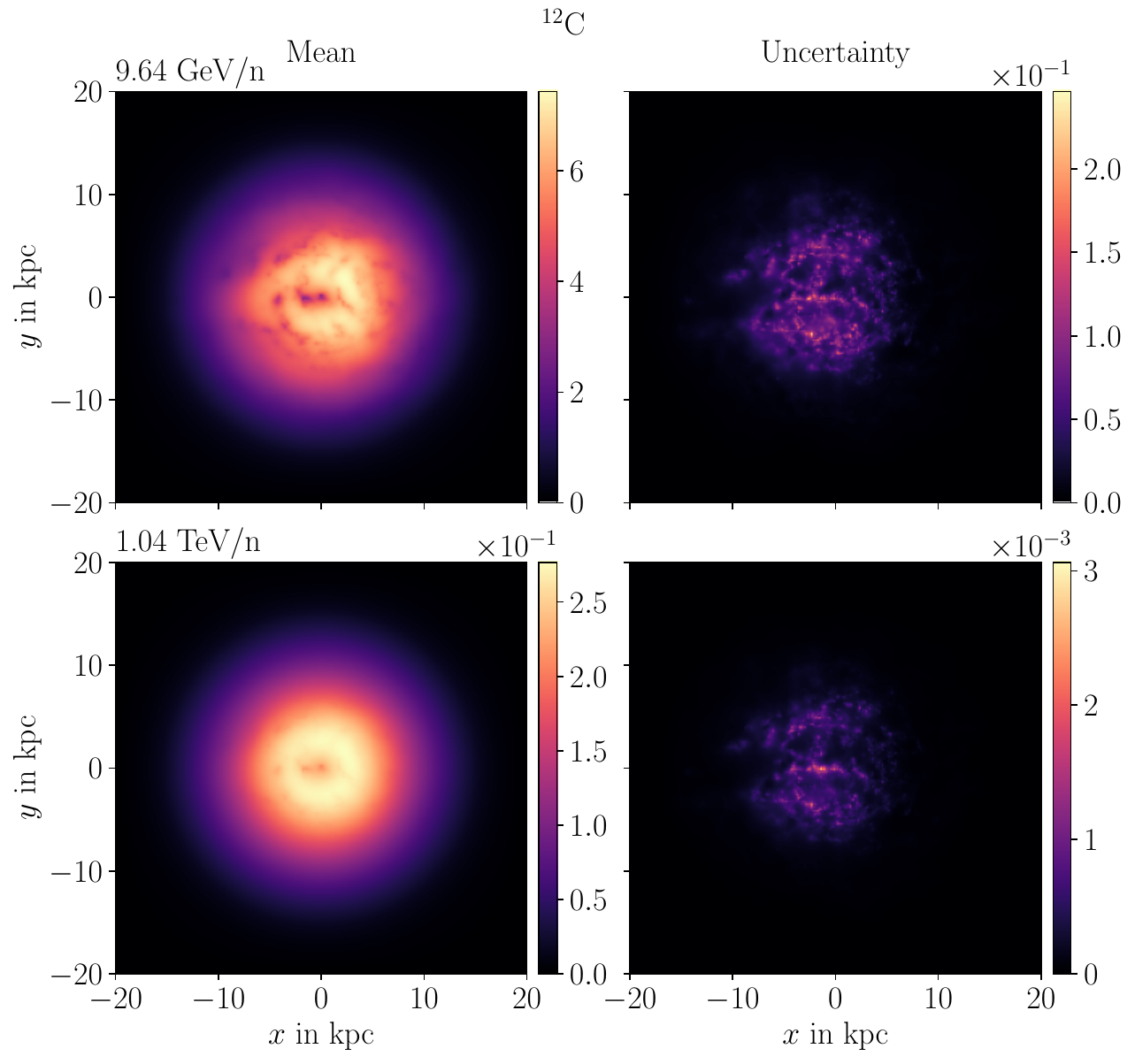}
    \includegraphics[scale=0.38]{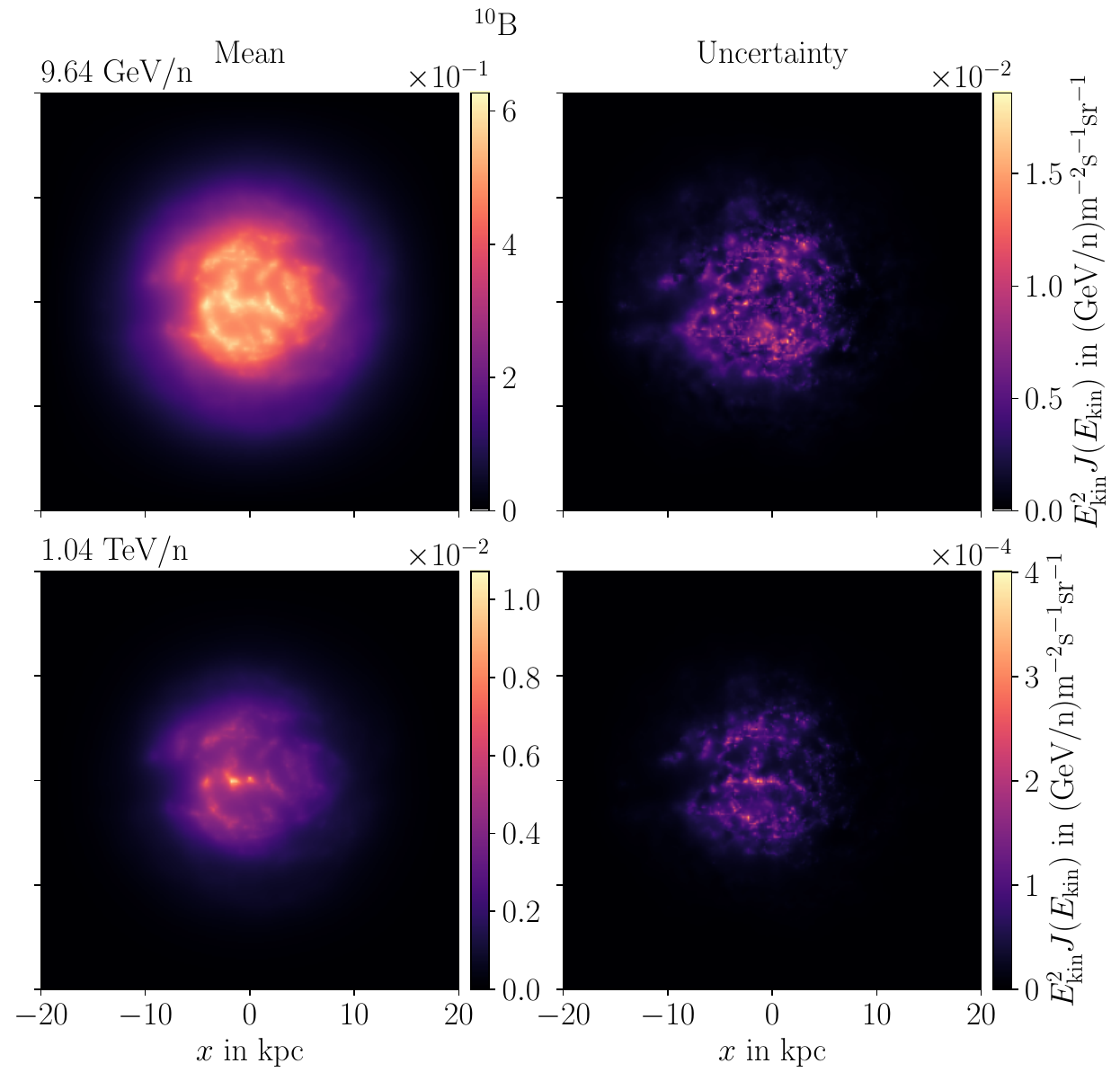}
	\caption{\textit{Left:} Galactic plane view of ${}^{12}$C mean distributions (left column) and uncertainties (right column) for the 3D model. \textit{Right:} Same for ${}^{10}$B. These quantities are presented for each nucleus type at 9.64\,GeV/n (top rows) and 1.04\,TeV/n (bottom rows).}
	\label{BoronCarbon}
\end{figure*}

\begin{figure*}
	\centering
	\includegraphics[scale=0.38]{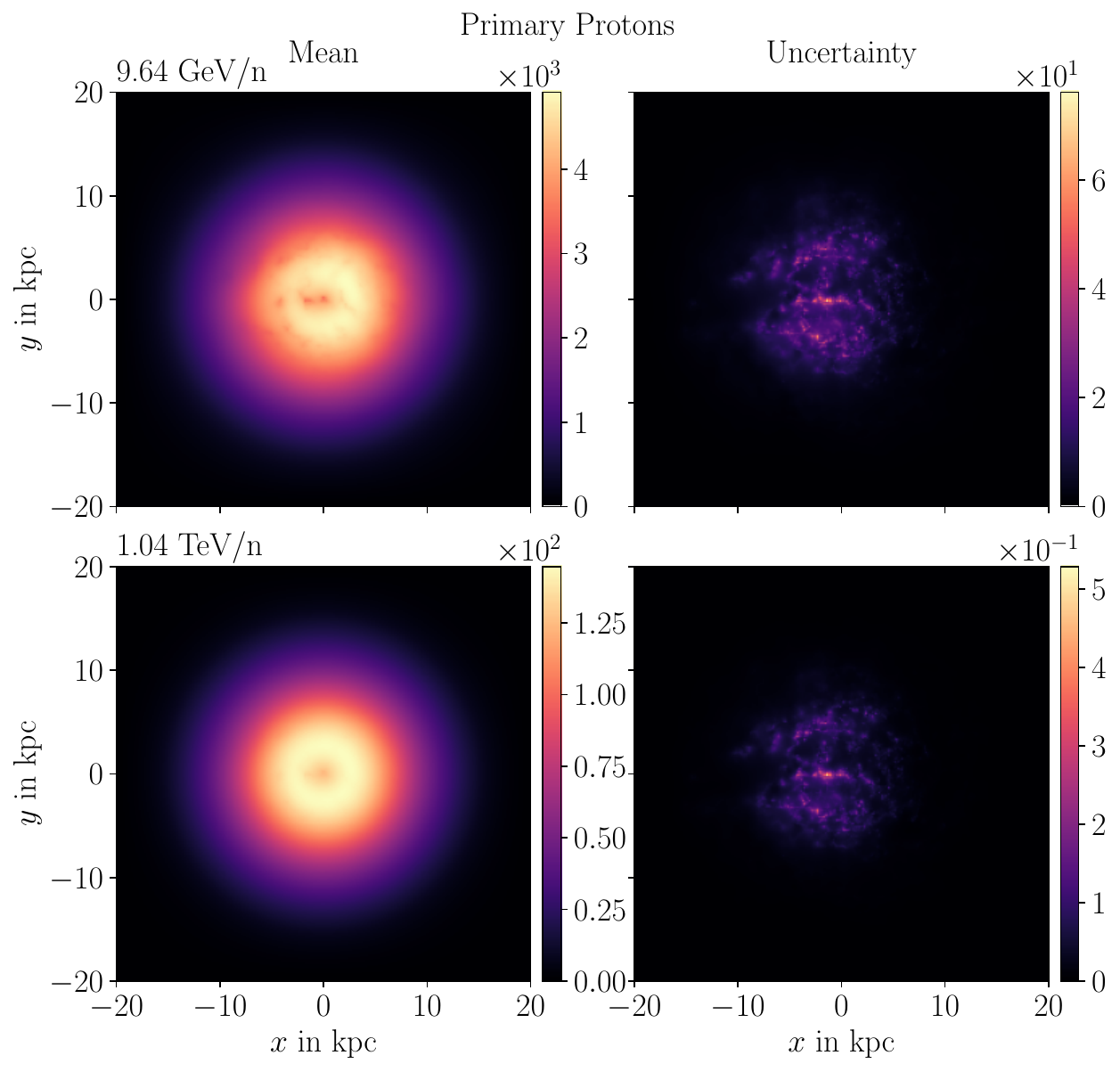}
    \includegraphics[scale=0.38]{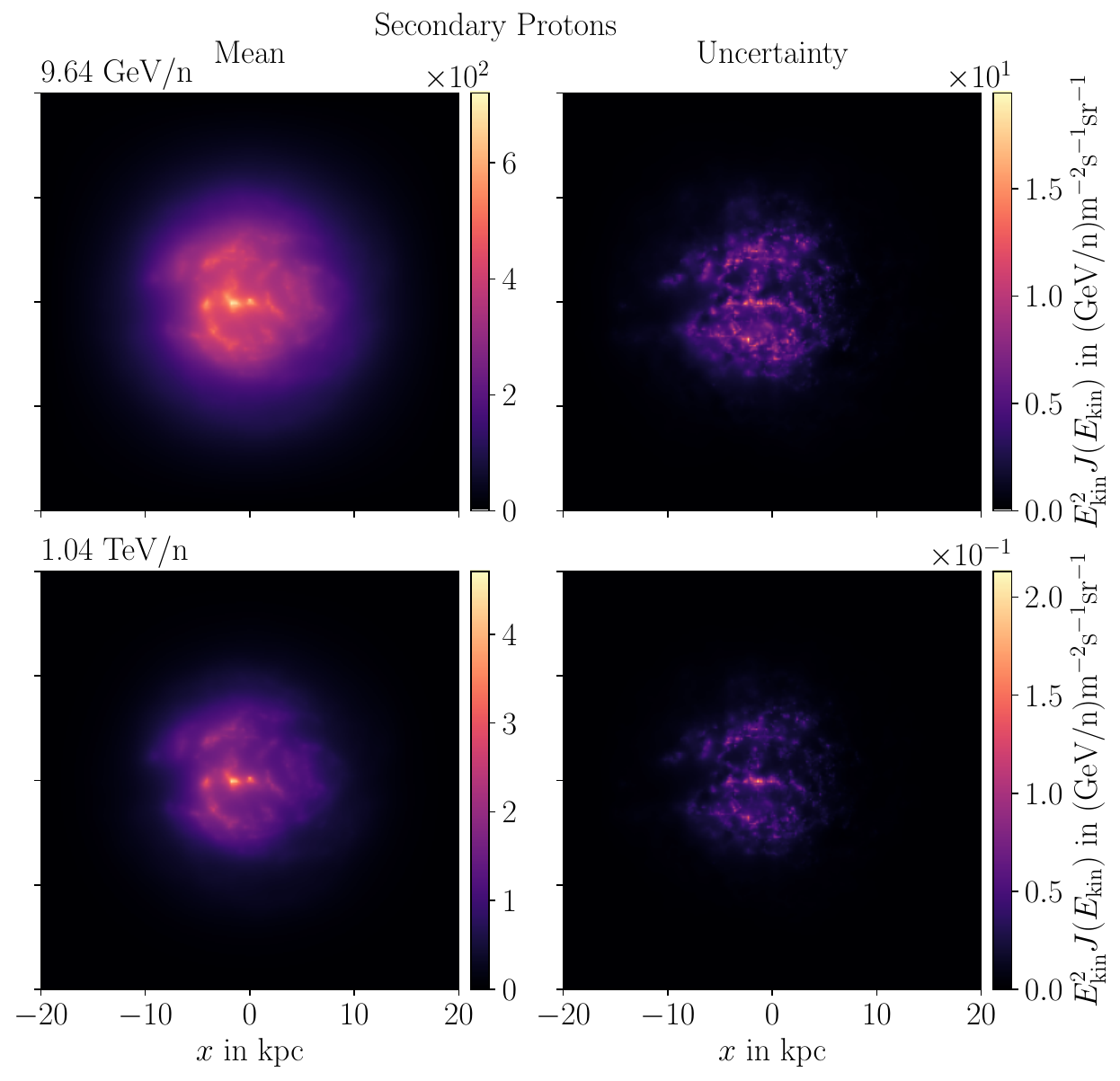}
	\caption{Like Fig \ref{BoronCarbon}, but for primary and secondary protons.}
	\label{Proton}
\end{figure*}

\begin{figure*}
	\centering
	\includegraphics[scale=0.38]{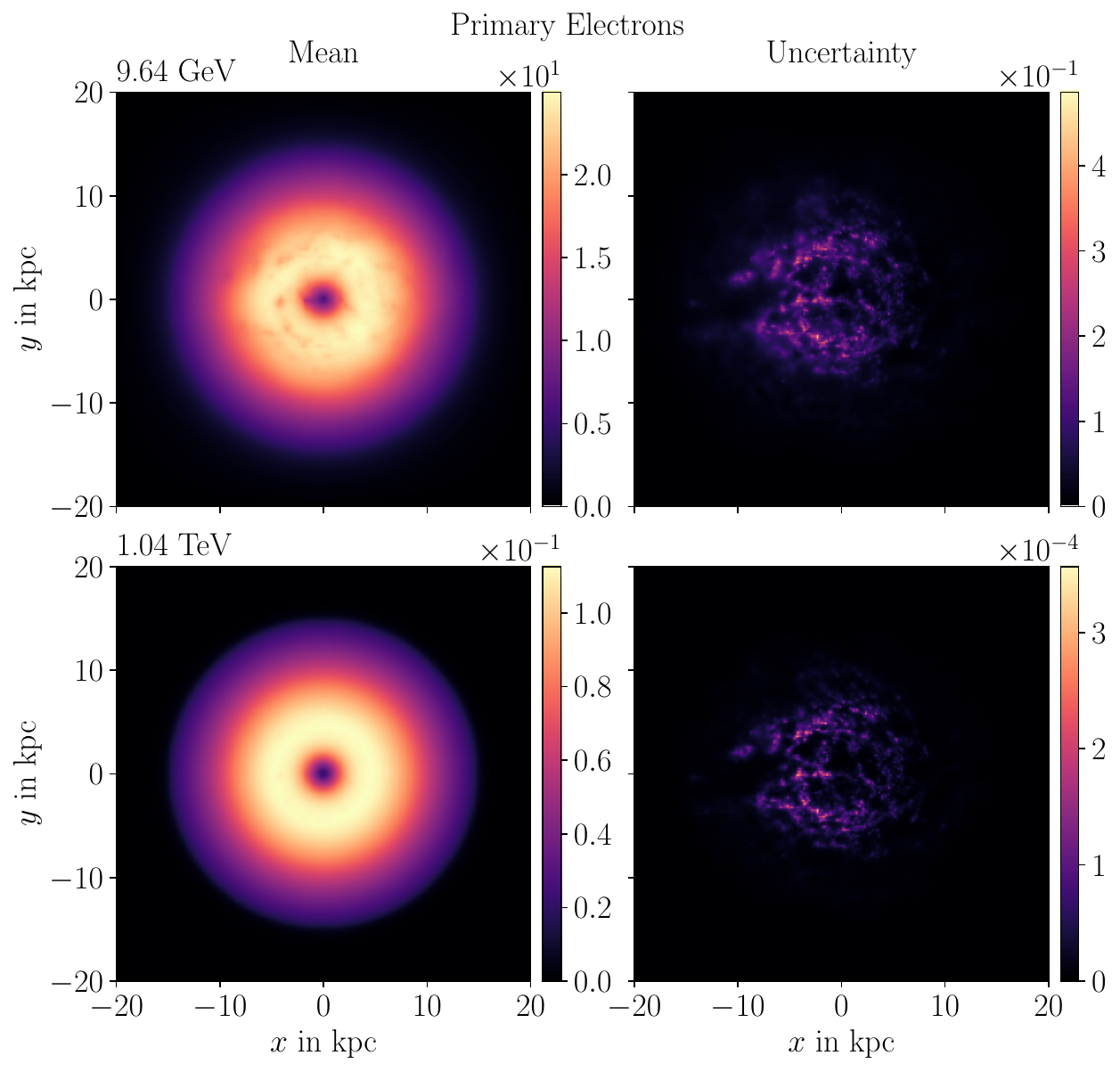}
    \includegraphics[scale=0.38]{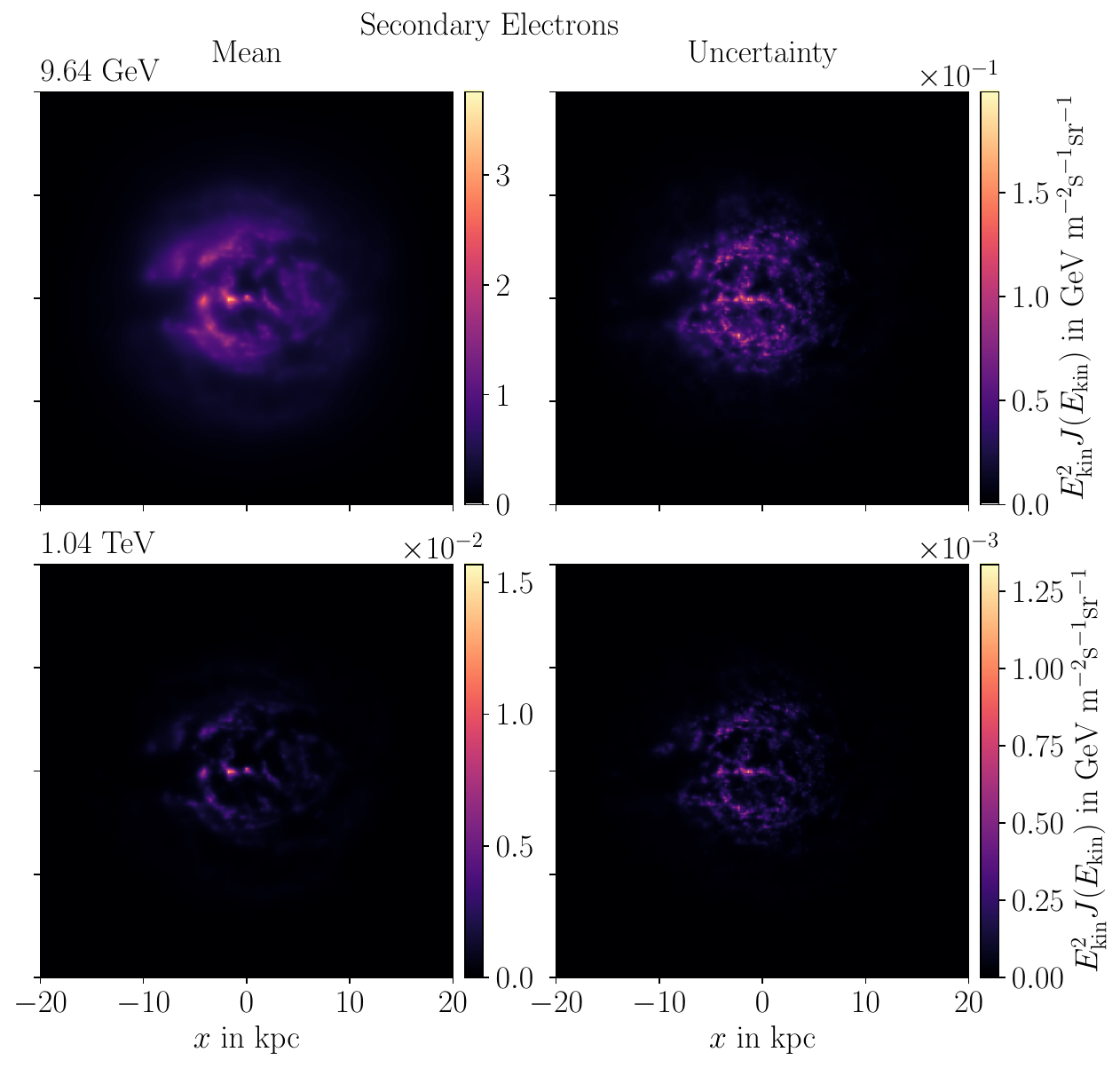}
	\caption{Like Fig \ref{BoronCarbon}, but for primary and secondary electrons.}
	\label{Electron}
\end{figure*}

All nuclei are subject to spallation in dense gas regions during transport, resulting in the production of secondary lighter nuclei at the expense of the interacting primary. As a consequence of this effect, simulations display under-populated regions that appear for primary nuclei where high density gas regions are located. In contrast, for secondaries, over-populated regions appear at high gas density regions. Hence, particles of mainly primary (secondary) nature are anti-correlated (correlated) with the distribution of gas. We illustrate this morphological impact of the 3D model on the distribution of different CR species,  particularly of leptons and heavy and light nuclei, due to the different dominant interaction processes with ISM matter at different energies. For this purpose, we look at the distributions at $\sim$10\,GeV/n and $\sim$1\,TeV/n of: Carbon 12 (${}^{12}$C), Boron 10 (${}^{10}$B), primary and secondary protons, and primary and secondary electrons.  (See Figs \ref{BoronCarbon}, \ref{Proton} and \ref{Electron}, respectively for the corresponding mean CR flux distributions and uncertainties in the Galactic plane).

We find that the corresponding CR uncertainties follow a profile similar to the gas uncertainty within 10 kpc from the GC, shown in Fig \ref{totprot}. For all species, the imprint of the morphology of the gas is evident for all energies. This imprint is stronger at lower energies, as seen at 9.64\,GeV/n in contrast to 1.04\,TeV/n. Particularly for secondaries (${}^{10}$B and secondary protons) at 9.64\,GeV/n, structures do not fully resemble the gas reconstruction profile. We note that for both primary and secondary species the relative uncertainties are much smaller for the CR distributions than for the gas reconstruction, decreasing further with increasing energy.

 We observe that under-populated regions of particles are present for all primary species and are more prominent at 9.64\,GeV/n, specially for ${}^{12}$C. Furthermore, the structure of such under-populated regions smoothens out with increasing energies until disappearing, a consequence of diffusion being the dominant transport process. In contrast, over-populated regions of particles are present at all energies for secondary species. At 1 TeV the source distribution of the CRs leaves a clear imprint for all species, where secondary CRs can be easily identified by their source distribution being related to the gas distribution.

 As a consequence of spallation and energy losses, we see that heavy species such as carbon and boron display more gas-induced structures at the GeV level in contrast to protons and electrons. For primary protons, losses due to interactions with gas are only significant at the densest regions, with inelastic losses and diffusion dominating at the chosen energies and at most locations on the plane \citep[see e.g.,][for the modeling of the relevant proton cross-section]{LCTan_1983}. For electrons, one important process is bremsstrahlung, which dominates electron losses in dense gas regions between $\sim$ 100 MeV and a few GeV, before becoming subdominant to IC and synchroton losses due to interactions with the interstellar radiation field (ISRF) \citep{PhysRevLett125051101}. Furthermore, due to the high energy losses at TeV energies, very high energy electrons do not travel far from their production sites -- being the ISM gas for secondary electrons -- \cite[see e.g.,][]{mertsch2020,evoli2021,thaler2023}, leading to a higher contrast and sharper structures in the electron distribution as energy increases, as seen at 1.04\,TeV in comparison to 9.64\,GeV for both primaries and secondaries.
 
\begin{figure}
	\centering
  	\includegraphics[scale=0.5]{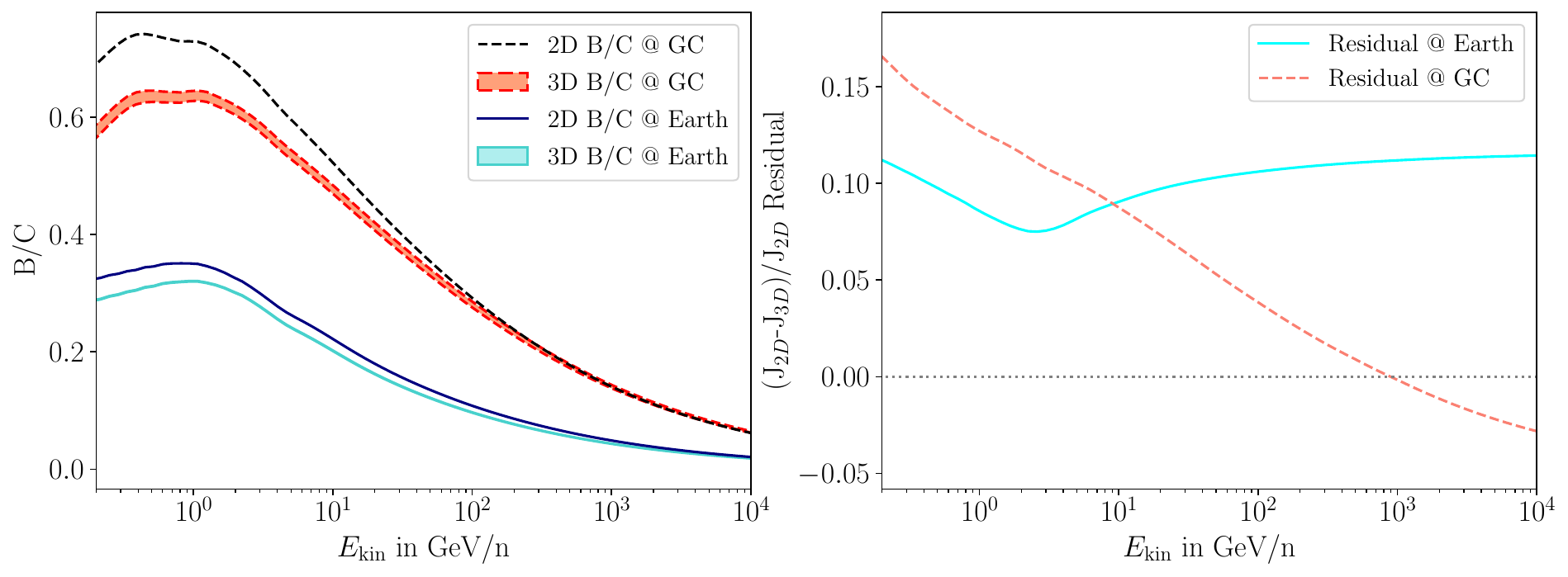}
	\caption{\textit{Left:} B/C ratio for the 2D and 3D models using the 3D model parameters for both. \textit{Right:} Fractional residual comparing the mean B/C ratio from the 3D model to the 2D model at both Earth and the GC. The collection of all simulations in the 3D model is represented by the bands covering the 1-$\sigma$ region of our samples. We show the B/C ratio at Earth (solid lines) and at the GC (dashed lines), where results at Earth are shown with a heliospheric modulation of 678.6\,MV.}
	\label{BC}
\end{figure}

We now investigate the impact of the gas distribution on CR spectra and the different impact in 2D and 3D models focusing on the B/C ratio. For comparison, we compute all the spectra from the 2D and 3D models shown herein using the same set of transport parameters given in Table \ref{tab:transportparameters} for the 3D model. We calculate these spectra at the Galactic center (GC) and at Earth's location, where the latter is fixed by the underlying velocity field model used to reconstruct the 3D gas density. For the BEG03 model, Earth is assumed to be located at 8.0\,kpc from the GC. We chose the GC region to study the impact of the distribution of the gas structures, since the average gas density within any distance $r$ from the GC is the same between the 2D and 3D models. For spectra at Earth we compare with CR data, where we use the force-field approximation to model heliospheric modulation. We take the modulation potential \citep[see e.g.,][for details on its estimation]{Ghelfi2017} from the AMS-02 measurements, as reported in the cosmic ray database (CRDB) \citep{Maurin2014,Maurin2020db,Maurin2023db}. To illustrate the interplay between primaries and secondaries in each model, in Fig \ref{BC} we show the B/C ratio at Earth and at the GC for both the 2D and 3D models. At both locations, as can be seen from the residual plot, we observe a higher B/C ratio in the 2D model than in the 3D model, up to 1 TeV at the GC and extending beyond 1 TeV at Earth. Flux differences at Earth are on average $\approx10.8\%$ at all energies. At the GC, such differences can be as large as $\approx17\%$ at 300\,MeV and decrease with increasing energy. At Earth, the average gas density differs between the 2D and 3D models -- with $\bar{n}_{\rm{2D}}\approx1.17\bar{n}_{\rm{3D}}$ as shown before -- leading to differences in the corresponding carbon and boron spectra. For both models, at the GC the local gas density is identical; thus, in this location, spallation and energy losses contribute in the same way to CR fluxes, and the resulting differences can only come as a direct consequence of the spatial distribution of gas structures itself, instead of the local average density. For individual CR species, we also note flux differences. At Earth, we find that CR fluxes below 10\,GeV can deviate up to 15\% for Carbon, 10\% for Boron, 5\% for protons, and 2.5\% for electrons. At the GC, these differences can double below 10
\,GeV for all species. We present the spectra of different individual species, including antiprotons, in \ref{appendix:spectra}.

\subsection{Correlation Structure}
\label{subsec:correlations}

\begin{figure}
	\centering
	\includegraphics[scale=0.7]{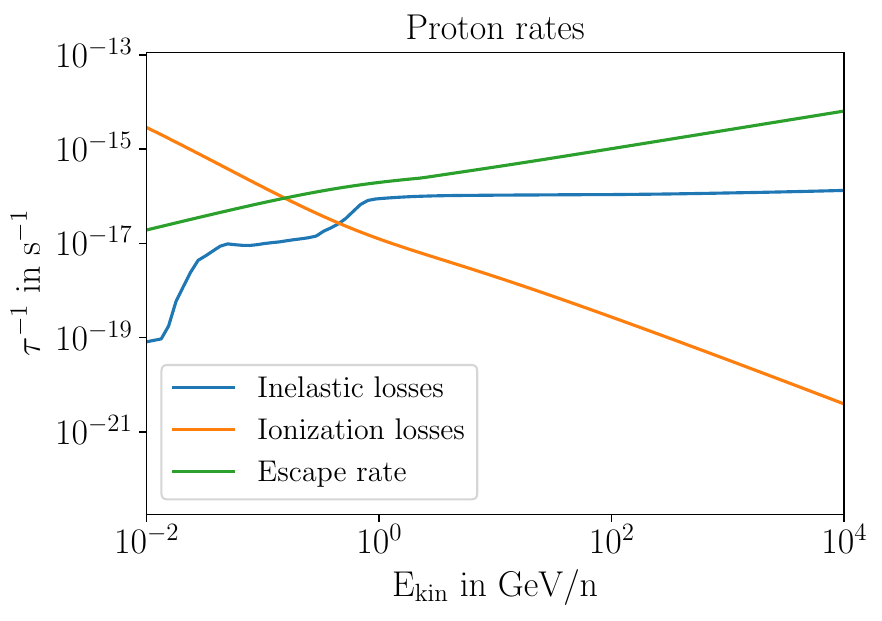}
	\caption{Interaction rates for protons. Ionization and inelastic losses are proportional to the local gas density. Here, $\bar{n}_{\rm gas}=0.05$ cm$^{-3}$ within 4\,kpc from Earth in the 3D model.}
	\label{timescales}
\end{figure}

\begin{figure*}
	\centering
	\includegraphics[width=\textwidth]{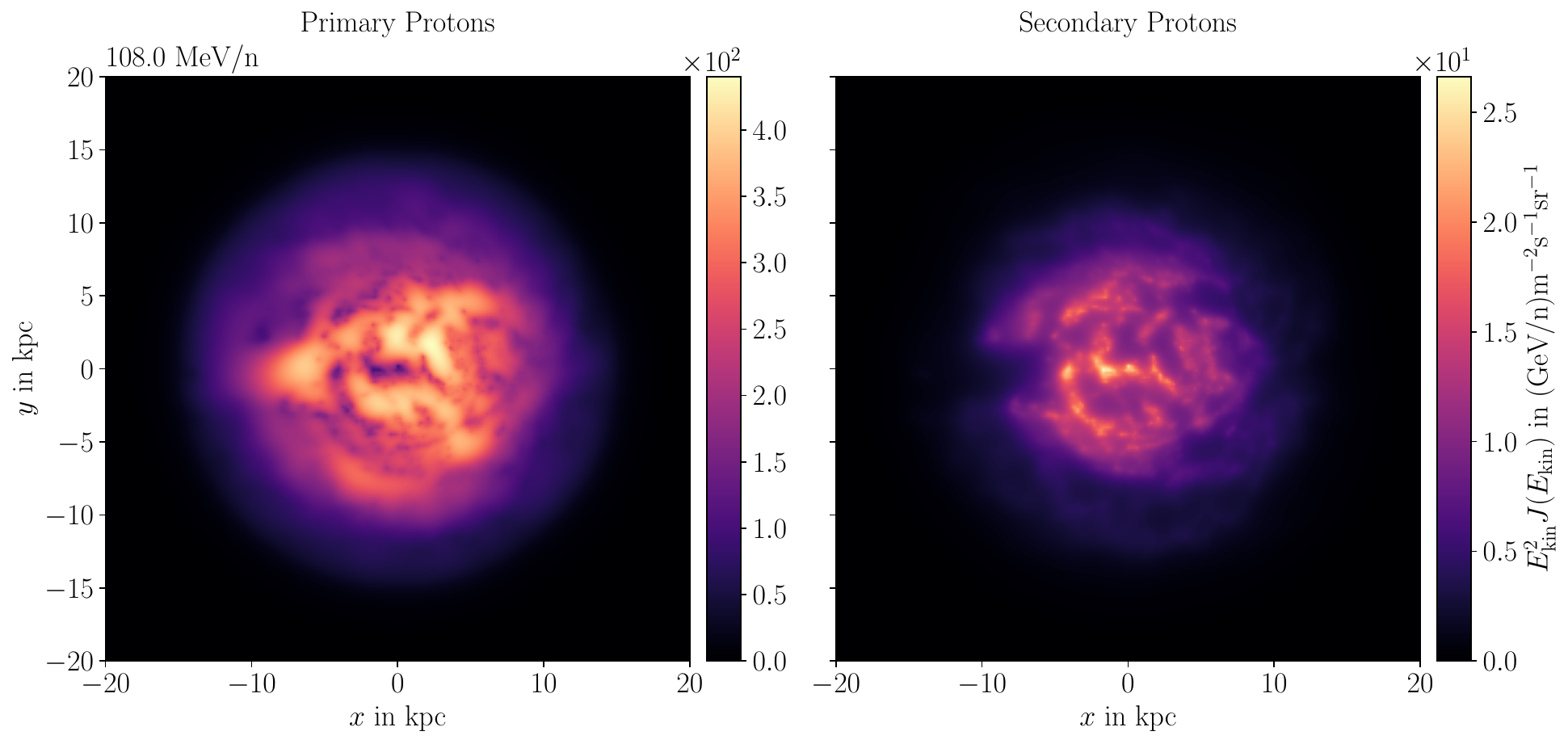}
	\caption{Distribution of primary (left) and secondary (right) protons on the Galactic plane at 108\,MeV/n.}
	\label{loweprotonslices}
\end{figure*}

\begin{figure*}
	\centering
	\includegraphics[width=\textwidth]{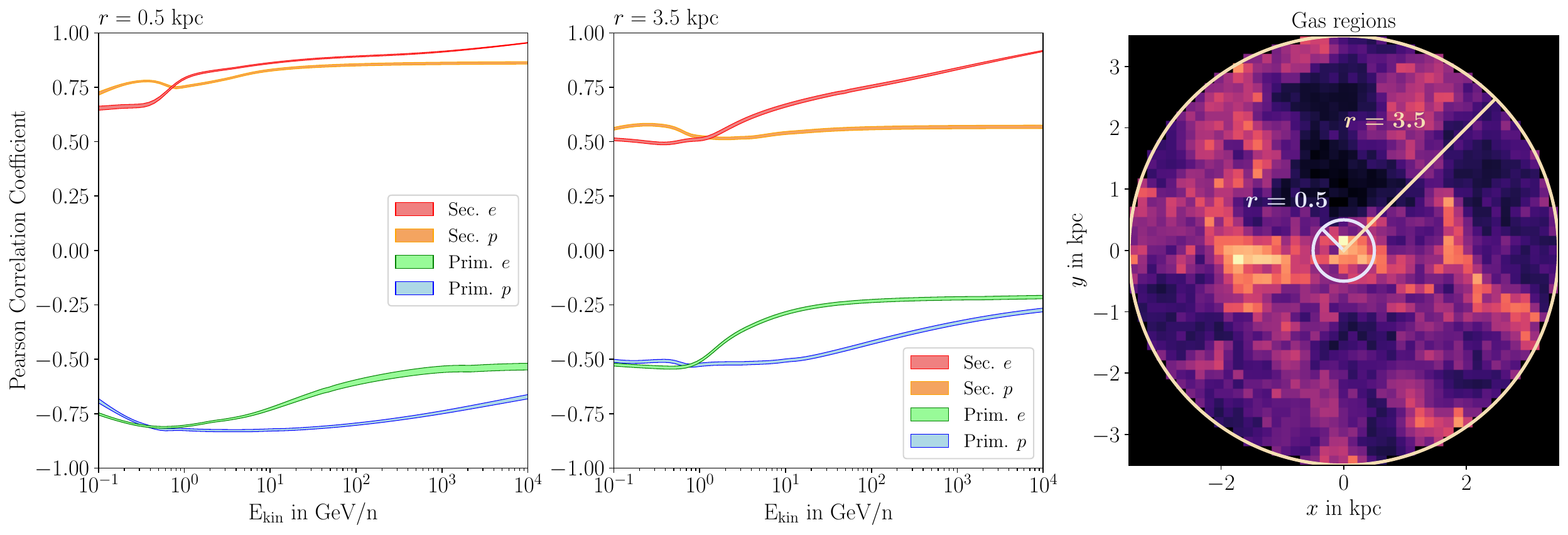}
	\caption{Pearson correlation coefficient at all simulated energies between electrons/protons and gas. We compute the coefficients for the Galactic plane regions within 0.5 and 3.5\,kpc from the GC. Each band covers the 1-$\sigma$ region over the 3D model simulations.}
	\label{correlationcoefficients}
\end{figure*}

Here, we display the correlation structure observed thus far in detail, by focusing on CR protons first to discuss how gas structures modulate their energy-dependent morphology. For this purpose, we study the proton interaction timescales at both the GC and Earth, the low-energy mean distribution of CR protons, and the correlation between gas and CR protons in the 3D model.
As previously stated,  and shown in \ref{appendix:spectra}, energy-dependent spatial structures follow from the spatial distribution of gas via the corresponding interactions of the different particle species. When such interactions contribute significantly to CR losses in comparison to the escape rate of CRs from the Galaxy, the resulting CR distribution exhibits local structures that follow the gas maps. 

In Fig \ref{timescales}, we show a comparison of different loss processes for protons, i.e. their inelastic, ionization, and escape loss rates averaged over a sphere with radius 4\,kpc around Earth. Here, inelastic losses are computed via their inelastic collisions and fragmentation interactions with ISM H and He, following the parametrization in \cite{LCTan_1983} for protons. Ionization losses, i.e. the energy loss rate due to the interaction of CRs that traverse neutral matter (i.e. ISM H and He), are adapted from \cite{MannheimSchlickeiser1994}. Finally, the escape rate is calculated as $1/\tau_{\rm{Esc}}=D/H^2$ with $H=4$\,kpc the halo height used in our model and $D$ the rigidity-dependent diffusion coefficient, using the values for the 3D model in Table \ref{tab:transportparameters}. Inelastic and ionization losses are directly proportional to the local gas density, while escape rates indirectly depend on the gas distribution by the change of $D$. Given a sufficiently high local gas density, these loss processes drive the presence of under-dense structures in CR primaries, compared to the escape rate of the corresponding CR.

To illustrate the impact of the gas distribution on the distribution of CR fluxes at lower energies, we show in Fig \ref{loweprotonslices} the Galactic plane distribution of primary and secondary protons at 108\,MeV/n. Although ionization rates dominate the low energy regime, primary protons are still subject to inelastic losses, which in turn drive the production of secondaries. We note that the ionization losses dominate for energies lower than $\approx 200$\,MeV/n, resulting in decreased primary proton fluxes in most dense gas regions within 10\,kpc of the GC. 

The under-dense features disappear at locations with sufficiently less dense gas as diffusion becomes dominant, leading to the smoothing of the distribution of primary protons visible in Fig \ref{Proton} for the distributions 9.64\,GeV/n and 1.04\,TeV/n. We note that the gas-induced structures that are still visible at 9.64\,GeV/n result due to inelastic losses being on a similar magnitude as diffusion. For secondaries, while the loss rates are the same, their injection occurs only at dense gas locations and thus their distribution is primarily determined by the gas maps.

To quantify this energy-dependent relationship between the structure of fluxes for primaries and secondaries, and that of the gas density, we calculate the Pearson correlation coefficient. For this, we use protons and electrons, and compare their distribution with the gas distribution inside 2 regions of radius $r=0.5$\,kpc and $r=3.5$\,kpc around the GC in the Galactic plane. The $0.5$\,kpc region considers the small over-dense patch of gas at the very center of the Galaxy, while the $r=3.5$\,kpc region takes into account local over- and under-dense gas clouds. The resulting energy-dependent correlation coefficients, along with the corresponding regions are shown in Fig \ref{correlationcoefficients}. We note the anti-correlation (correlation) between primaries (secondaries) and gas density in both regions at all energies. For very-high-energy secondary electrons (i.e. $E_{\mathrm{kin}}=>100$\,TeV/n) the correlation is very close to 1, reflecting an almost linear dependence between gas and secondary electrons as a consequence of their increasingly strong confinement to their source regions (i.e. the high density regions) due to their high energy losses. We observe that with increasing energies, the absolute value of the correlation coefficient between secondaries and gas is bigger than the one between primaries and gas. This comes as a consequence of the anti-correlation with primaries being driven by gas-dependent losses. On the $r=3.5$\,kpc region, the correlation coefficient for secondary protons becomes considerably smaller, since they can travel larger distances than electrons due to lower energy losses.

\subsection{Gamma-ray emission}
\label{subsec:gammaemission}

\begin{figure}
	\centering
	\includegraphics[scale=0.6]{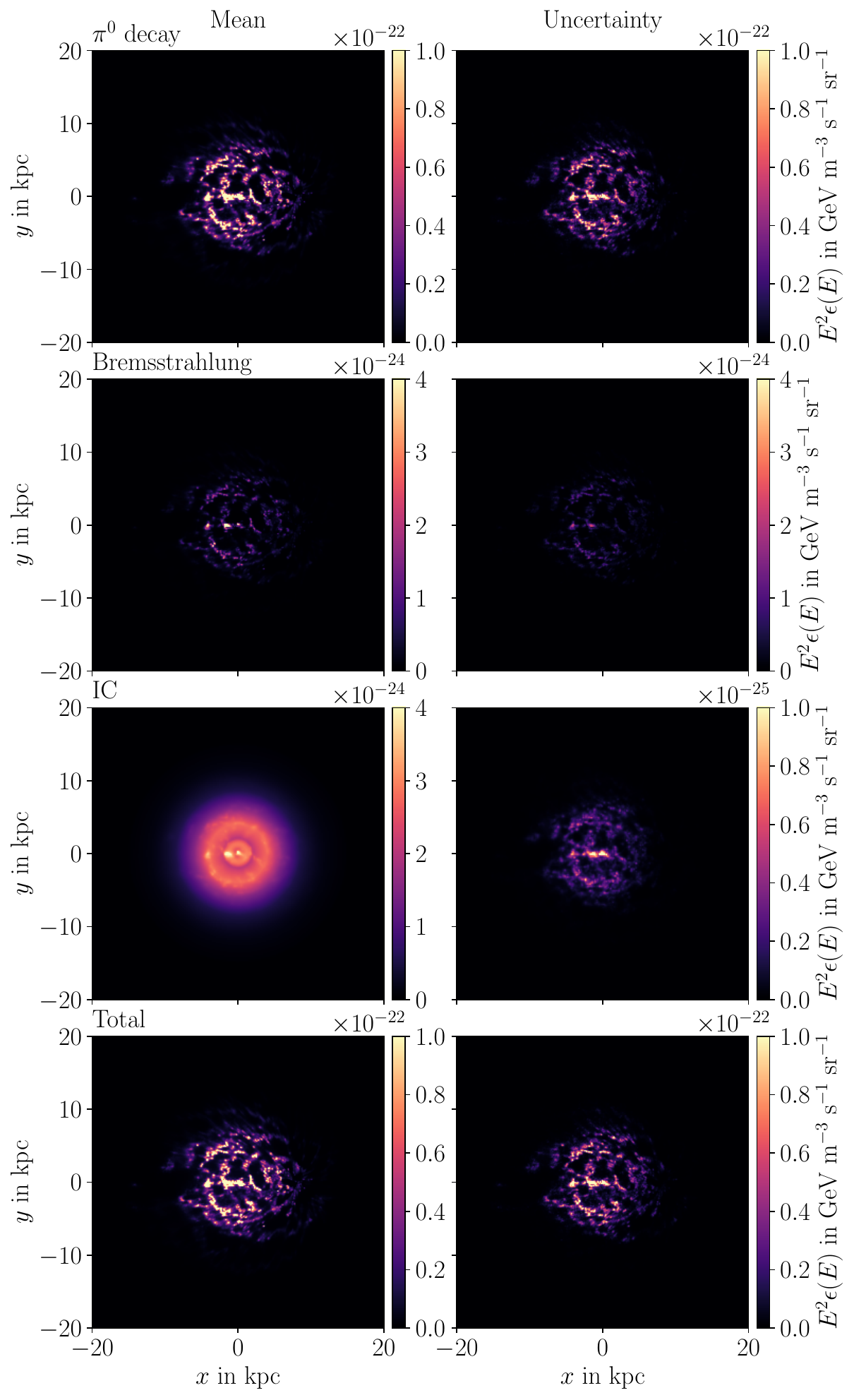}
	\caption{Mean distribution (left) and standard deviation (right) of the $\gamma$-ray emission channels in the Galactic plane at 1\,TeV. Emission channels from top to bottom: $\pi^0$-decay, bremsstrahlung, IC, and Total. Bremsstrahlung and IC, are shown using the same color scale for comparison.}
	\label{Gamma Emission}
\end{figure}

Here, we show our $\gamma$-ray emission results, which are based only on the 3D model. Like for the CR transport results, we check the morphological impact of the 3D model on the individual emission channels. We identify their relation to the distribution of gas by studying the mean $\gamma$-ray spectra at different over-/underdense gas regions in the Galactic plane. 

In Fig \ref{Gamma Emission}, we show the mean distributions and uncertainties of the simulated local emissivities for $\pi^0$-decay, bremsstrahlung, and IC $\gamma$-ray emission, as well as for the total $\gamma$-ray emission (including IC). We show these results in the Galactic plane at 1\,TeV. The profile of the total $\gamma$-ray emission is directly correlated with dense gas regions as a consequence of $\pi^0$-decay being the dominant emission channel. $\pi^0$-decay dominates by 2 orders of magnitude at this energy. %Unlike the CR transport case, $\pi^0$-decay and bremsstrahlung emission scale linearly with gas density, and the corresponding uncertainty is similar to the one of the gas reconstruction, in particular, at locations of high H$_2$ density. 
Furthermore, $\pi^0$-decay and bremsstrahlung emission exhibit uncertainties of magnitudes comparable to their corresponding mean distribution.

In contrast to the other channels, IC emission follows from the interaction between CR electrons and the ISRF, where the latter is introduced in the simulations by a 2D model, as discussed in section \ref{sec:Setup}. In this case, structures related to the 3D gas maps are faint. Unlike the other emission channels, IC emission does not scale linearly with the gas density. As a result, features on the emissivities only appear due to the existing structures in the distribution of CR electrons, which explains the uncertainties being negligible, as seen for the CR fluxes.

\begin{figure*}
	\centering
	\includegraphics[scale=0.38]{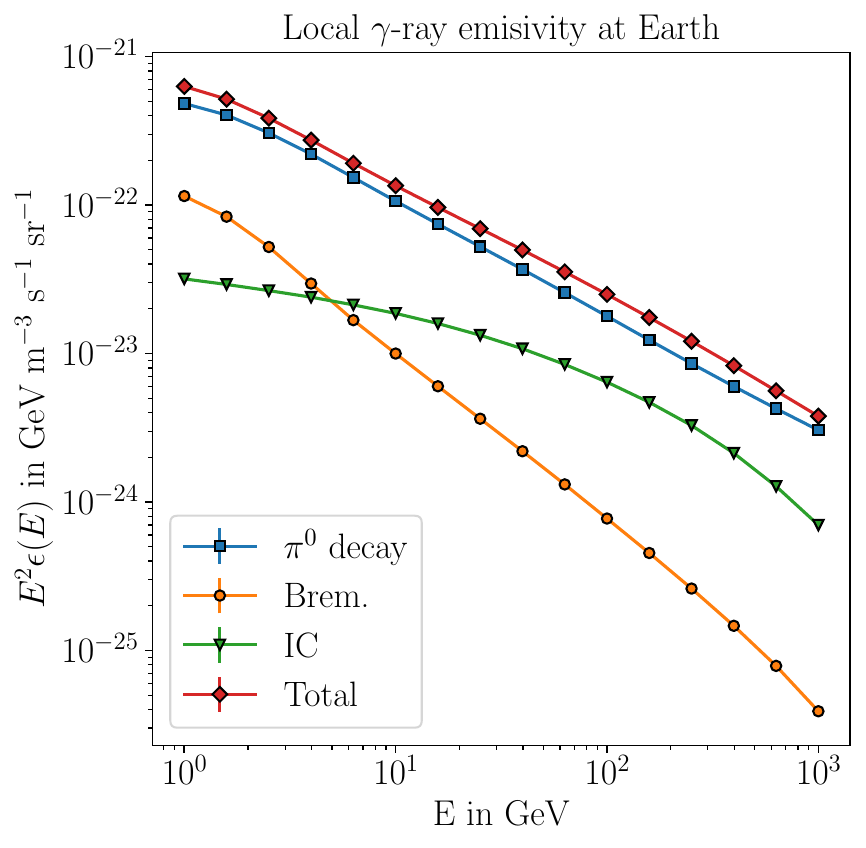}
	\includegraphics[scale=0.38]{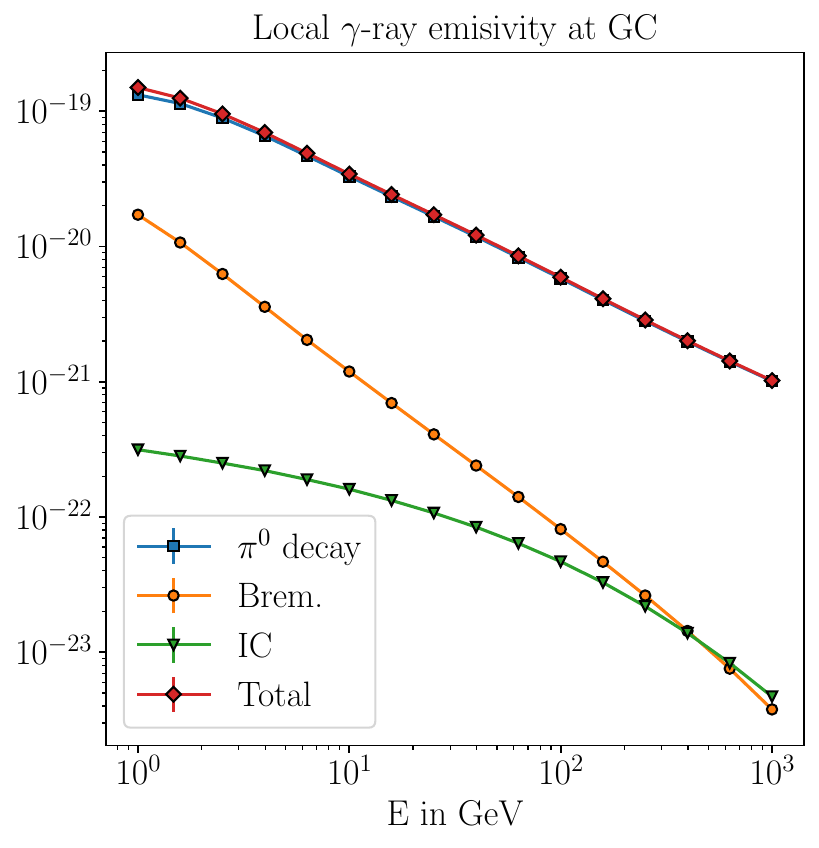}
	\includegraphics[scale=0.38]{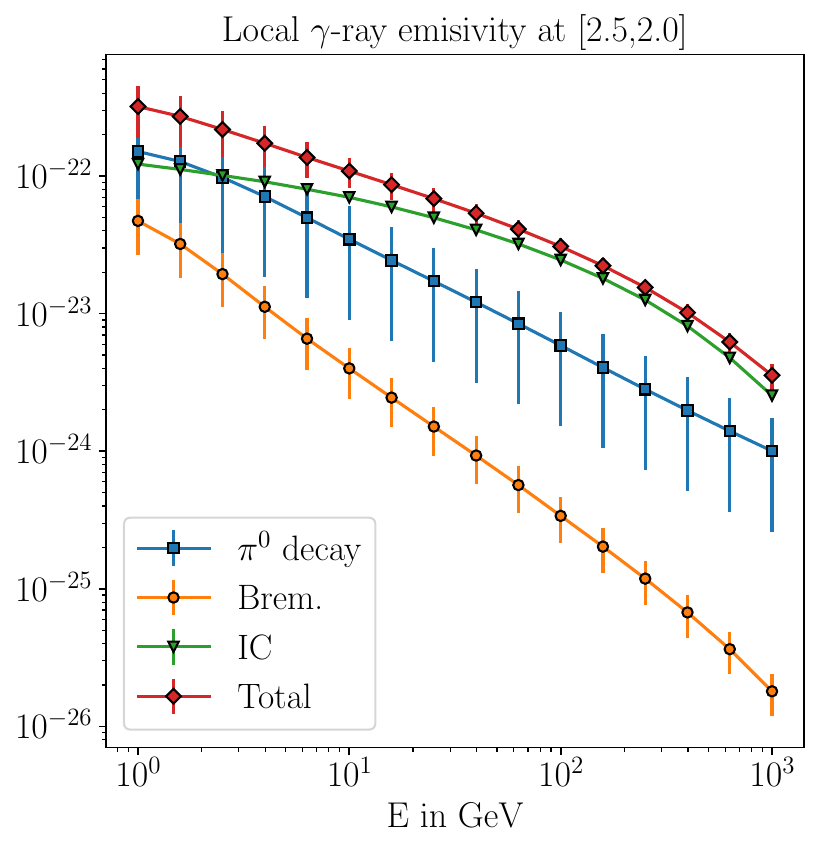}
	\caption{Total and individual  local $\gamma$-ray spectra channels simulated at Earth's location (left), at the GC (middle) and in the under-dense region located at $[2.5,2,0]$.}
	\label{Gamma Emission Spectra LOC}
\end{figure*}

In Fig \ref{Gamma Emission Spectra LOC} we show the $\gamma$-ray spectra for the total and individual $\gamma$-ray emission channels, where we find a different slope change at the location ($x=2.5$,$y=2.0$)\,kpc compared to the GC that is related to a local dominantion of IC emission. As expected, the emission is mostly proportional to the gas density, being highest at the GC. The uncertainties observed in $\gamma$-ray spectra are much larger than for the CR spectra and correlate directly to regions with high gas density, where the uncertainties of the gas are also highest. The local dominance of IC emission is a consequence of the joint effect of the ISRF being high and a locally low gas density. At Earth, $\pi^0$-decay emission is dominant with IC emission becoming comparable in the 100-600\,GeV energy range. At the GC, the gas density is so high bremsstrahlung emission dominates over IC emission. In our gas model we find a low uncertainty at the GC, which is a consequence of the underlying velocity field model and leads to negligible uncertainties of $\gamma$-ray emission at this exact location. 

\begin{figure}
	\centering
	\includegraphics[width=\columnwidth]{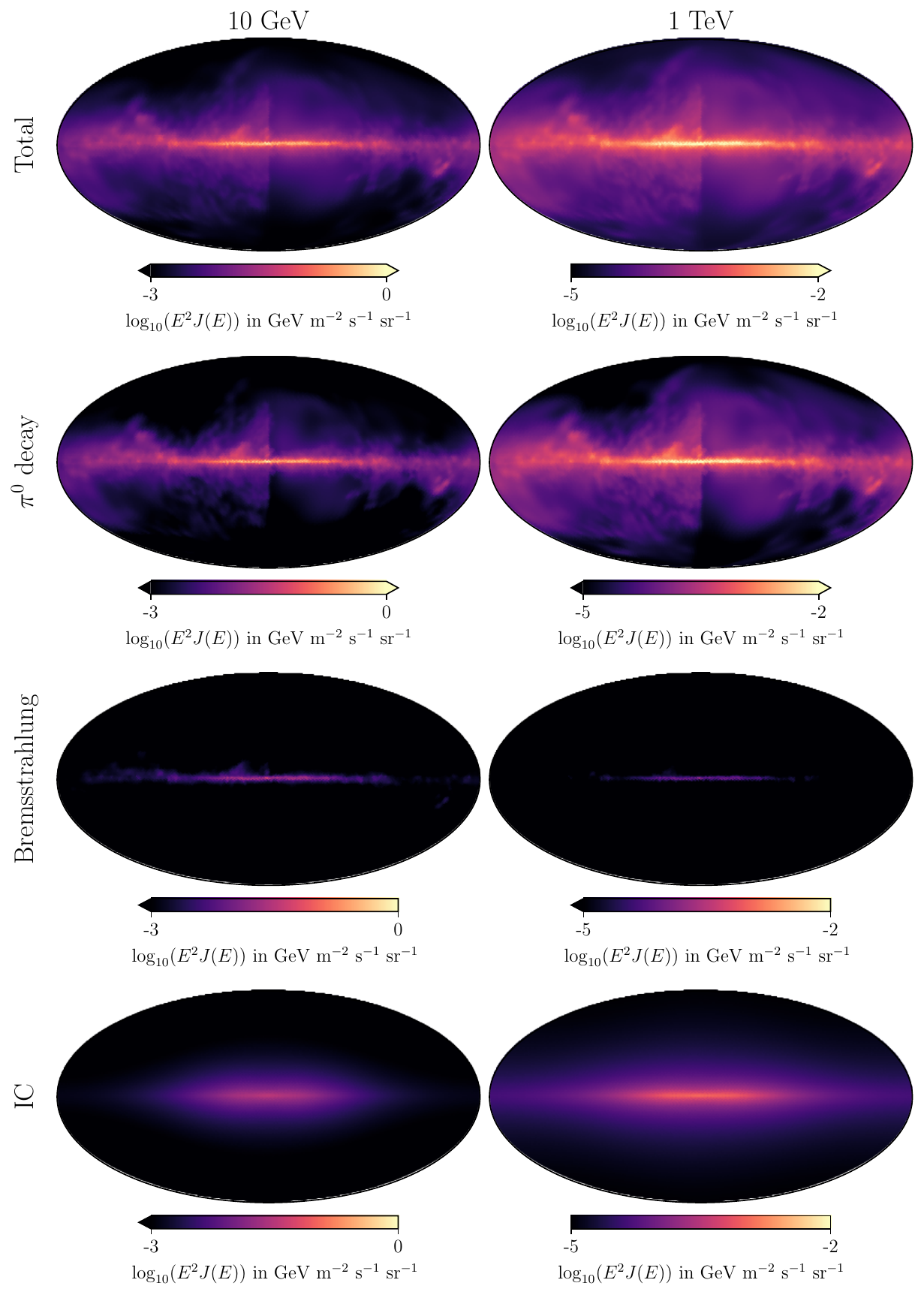}
	\caption{All sky $\gamma$-ray emission for 3D model at 10\,GeV (left) and 1\,TeV(right). From top to bottom: Total, $\pi^0$-decay, bremsstrahlung, IC emission. Maps are shown for 10\,GeV (left) and 1\,TeV (right) energies.}
	\label{LOS Gamma Emission}
\end{figure}

\begin{figure}
	\centering
	\includegraphics[width=\columnwidth]{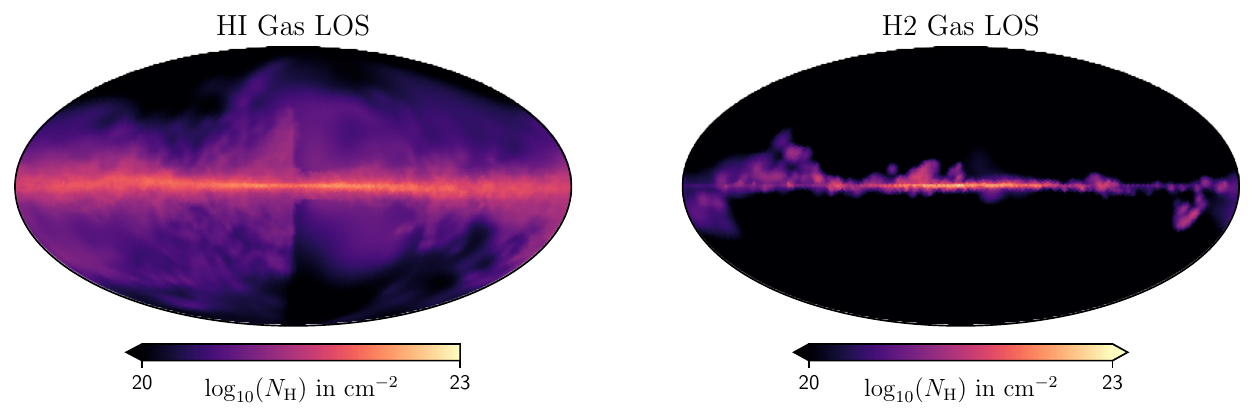}
	\caption{All sky map for the individual HI (left) and H$_2$ (right) gas reconstructions from the BEG03 model.}
	\label{LOS gas}
\end{figure}

In Fig \ref{LOS Gamma Emission}, we show the LOS-integrated  emission maps for all channels at 10\,GeV and 1\,TeV. We note that all major structures in the total emission follow from the dominating $\pi^0$-decay emission. For bremsstrahlung emission, contributions are concentrated in the Galactic plane where the gas density is highest. Similar to its corresponding local emissivity, the LOS-integrated IC emission is not strongly correlated with the gas density and does not exhibit gas structures on the sky map. Furthermore, IC emission extends to higher latitudes at 1\,TeV than at 10\,GeV. For reference, in Fig \ref{LOS gas} we show the individual HI and H$_2$ all-sky maps. In the case of bremsstrahlung, all structures are mostly determined by the distribution of H$_2$, as this contributes the highest density in the Galactic plane, where CR $e^\pm$ are densest. In contrast, $\pi^0$-decay emission exhibits structures from both H$_2$ and HI, where the high-latitude emission clearly follows the distribution of HI gas.

As the original gas reconstructions are defined on a Cartesian grid, the LOS projection is affected by the different angular resolution between nearby and distant gas structures. This effect is most obvious at high longitudes, where structures on the LOS projections are blurred out. At small longitudes the discrepancies between nearby and distant structures are less blurred. However, this also highlights prominent reconstructions artifacts such as the sharp jump seen at the GC.

\begin{figure*}
	\centering
    \includegraphics[scale=0.38]{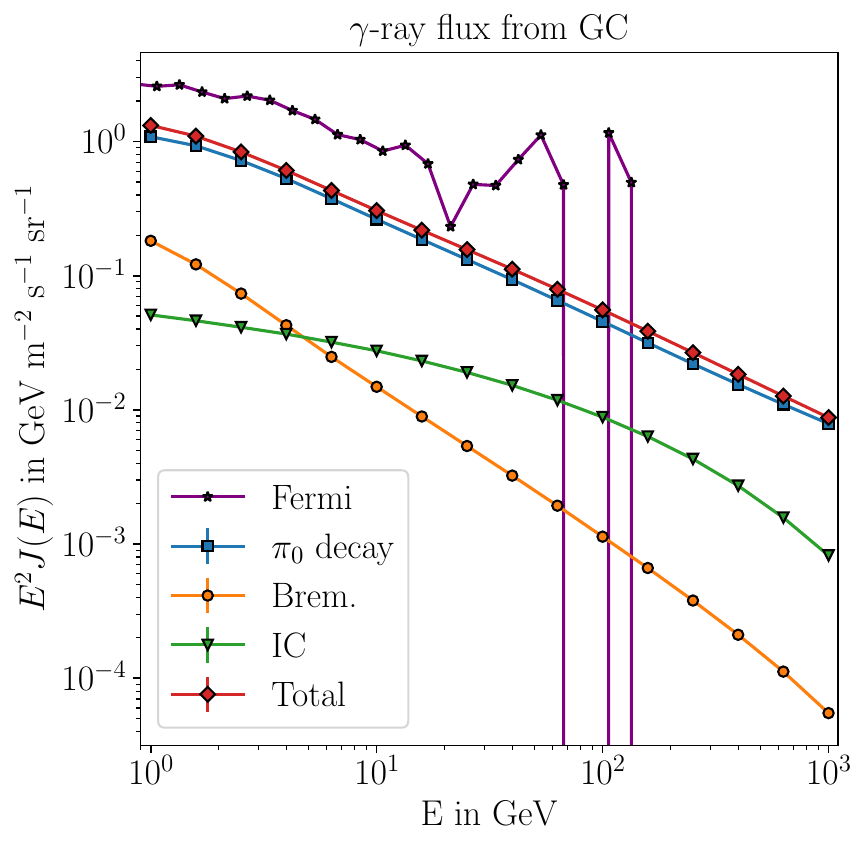}
	\includegraphics[scale=0.38]{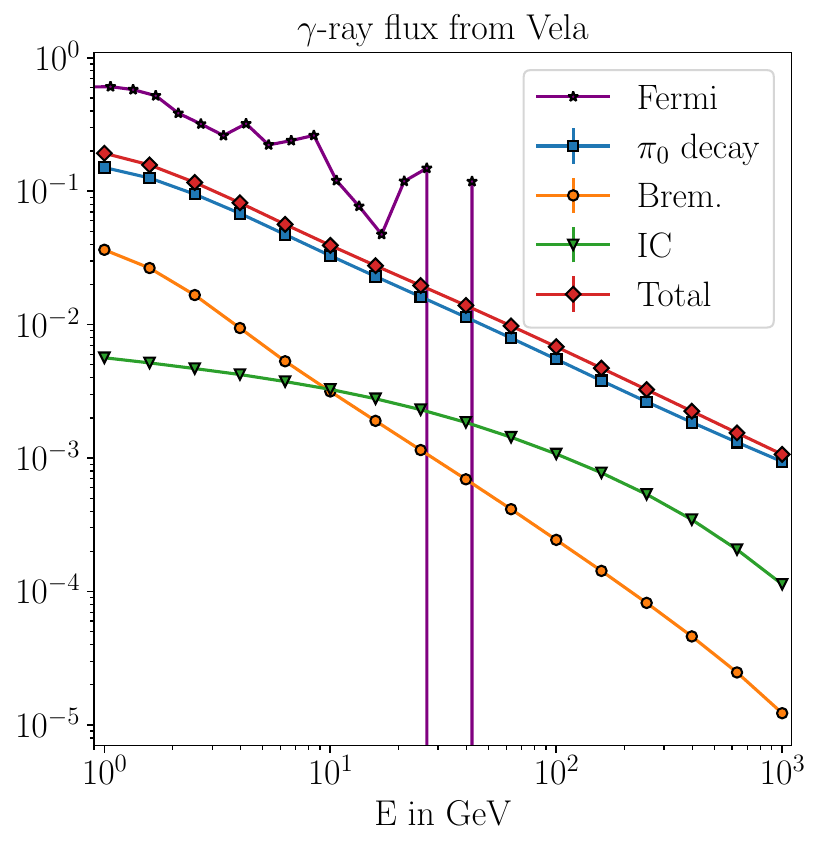}
	\includegraphics[scale=0.38]{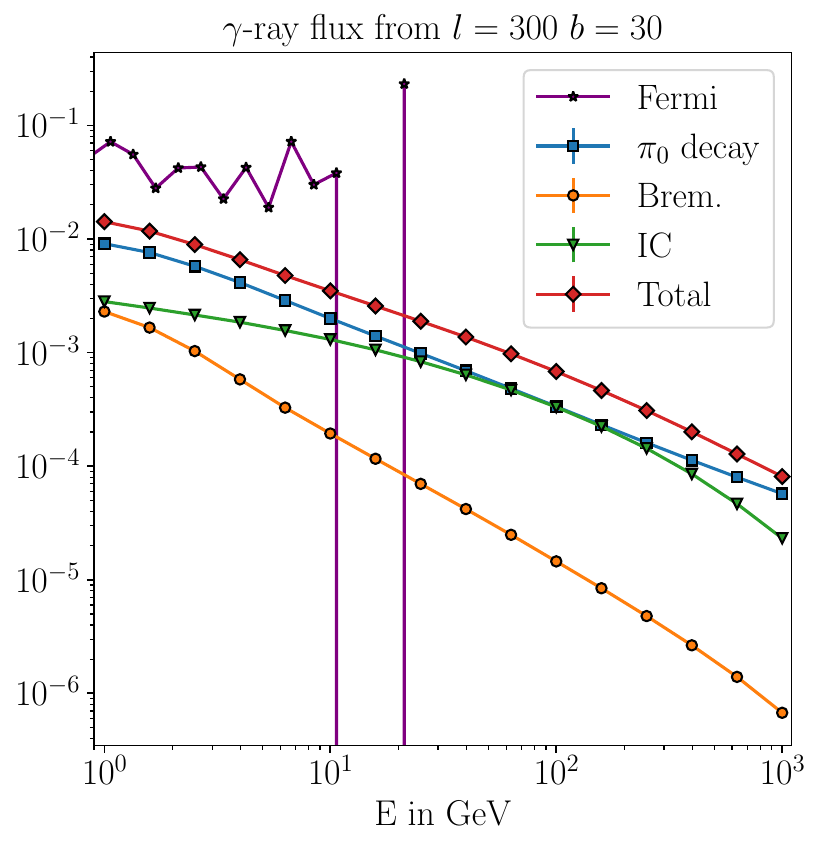}
	\caption{Total and individual $\gamma$-ray spectra channels observed from Earth (left) Vela (middle) and at $(l,b)=(\ang{300},\ang{30})$.  
    Spectra follow by averaging the contribution within a disk of radius \ang{1} around the relevant location. Fermi-LAT data (extracted as in \citet{fermigamma}) from each corresponding region in the sky is shown for comparison.}
	\label{GammaSpectraLosLoc}
\end{figure*}

In Fig \ref{GammaSpectraLosLoc} we show LOS-integrated spectra for the individual $\gamma$-ray spectra emission channels from the GC, Vela and the under-dense region on the sky at $(l,b)=(\ang{300},\ang{30})$. Our predictions are compared to the Fermi-LAT PASS8 data -- including point sources, unresolved sources, and the isotropic extragalactic component -- extracted as in \citet{fermigamma}. Like for the local emissivities in Fig \ref{Gamma Emission Spectra LOC}, at the under-dense region, IC emission itself is comparable to $\pi^0$-decay emission. We note that the uncertainties become negligible, as they are defined for each voxel in the 3D gas distribution and have little effect after integrating along the LOS. Furthermore, we find that the spectral slope of the $\gamma$-ray spectra from dense regions like Aquila, the Cepheus Flare and the under-dense region at $(l,b)=(\ang{300},\ang{30})$ varies slightly, due to the contribution of IC emission. We see that despite the local bremsstrahlung emission boost at the GC, the integrated contribution from this direction does not dominate over IC emission. In \ref{appendix:gammaemission} we show the total $\gamma$-ray emission spectra at several additional locations, both for local emissivities and integrated along the LOS.

LOS-integrated $\gamma$-ray spectra can also be calculated using the CR transport results of the 2D model, using the 3D gas as a target. This approach follows earlier methods employed for modeling Galactic diffuse emission, in which two gas models are used: one to simulate CR propagation; and a different one, with an appropriate column density, is used as a target for $\gamma$-ray emission. In this case, deviations of up to 5\% in most regions occur between 2D and 3D models, and in over-dense regions such as the vicinity of the GC such deviations can increase to $\approx$13.5\%. See \ref{appendix:Gamma-model-diffs} for further details on how the choice of the target gas distribution influences $\gamma$-ray emission.

%% file: Sections/Conclusion.tex
\section{Summary and conclusions}
\label{Summary and conclusions}

We studied the impact of new 3D reconstructions of Galactic HI and H$_2$ on our understanding of the CR transport and $\gamma$-ray emission. For this, we prepared a 2D and a 3D model, where the 2D model corresponds to an azimuthally averaged version of the new gas maps interpolated to the simulation grid used, while the 3D model directly interpolates the new gas maps onto the simulation grid. Then, we solved the CR transport equation using PICARD for isotopes up to silicon. Lastly, we used the corresponding transport results for the computation of Galactic diffuse $\gamma$-ray emission. For our simulations, we used 20 possible realizations of the new gas reconstruction to determine the impact of the uncertainties of the gas distribution at all stages of the simulations.

For CR transport, we found that, to account for a 3D structured gas distribution when fitting to B/C ratio data, the diffusion coefficient $D_{xx}$ and Alfvén velocity $v_{\text{A}}$ needed to be decreased in comparison to the 2D model, effectively reducing the diffusive strength. Furthermore, the rigidity dependence of the diffusion coefficient needed to be adjusted. We found that, with a more realistic gas distribution, the adjustment of these parameters was sufficient to achieve a fit to B/C ratio data with a $\chi^2$ value of 1.47. Likewise, our predictions for antiprotons led to a $\chi^2$ value of 1.89 when comparing to data.

Next, we investigated the gas-induced structures in the distribution of different CR species, namely Boron, Carbon, protons and electrons. We observed that structures are gas-density, energy and CR-species dependent, where the different fragmentation and energy loss processes in the ISM lead to different imprints of gas in both primary and secondary species. We found a clear anti-correlation between gas density and CR fluxes for primaries at energies where diffusion does not dominate. In contrast, for secondaries there is a direct correlation between gas density and CR fluxes at all energies. This can clearly be seen by the energy-dependent Pearson correlation coefficient between protons/electrons and gas. We found that the uncertainties propagated from the uncertainties of the published gas distributions were negligible for the final CR distributions.

When comparing the results of the 3D model with those of the 2D model using the same transport parameters, we found that at Earth, CR fluxes deviate noticeably at energies below 10\,GeV, up to 15\% for Carbon, 10\% for Boron, 5\% for protons, 2.5\% for electrons. Such deviations doubled at the GC, where the only difference between 2D and 3D models results from the local distribution of gas, since the average gas density is identical in both cases. Hence, the local distribution of gas structures in dense gas regions appears to significantly modulate the distribution of CRs. Furthermore, species of mainly secondary nature (Boron, anti-protons, TeV electrons where secondaries dominate) exhibited deviations across all energies, being 10\% in the energy range studied for antiprotons at Earth and Boron at the GC. However, the deviations observed were in general more significant for nuclei, as seen for Carbon below 10\,GeV.

Finally, we studied $\gamma$-ray emission produced from the simulated CR species both locally in the Galactic plane via the LOS integrals. In the Galactic plane, we observed a direct correlation between the total $\gamma$-ray emission and dense gas regions. In general, $\pi^0$-decay and bremsstrahlung emission are highly sensitive to fluctuations in the gas density. Therefore, the uncertainties in the gas distributions used for their computation are directly translated to the computed emission, as we saw in the corresponding 3D model results. In locations of high gas density, the $\pi^0$-decay channel always remained dominant, with bremsstrahlung emission being boosted and dominating over IC emission in such regions, like the GC. In low gas-density regions, IC became dominant, conditional to a locally high ISRF. From the LOS-integrated emission, the structure of the total emission is fully determined by the gas maps. For the individual emission channels, gas structures on IC are smoothed out.

We performed an equal-setup comparison between 3D and 2D gas maps in our models, which allowed us to quantify the contribution from the structured nature of dense gas clouds to CR fluxes and $\gamma$-ray emissivities. Thanks to the multiple gas realizations available from the reconstructions, we have for the first time calculated how uncertainties in such a gas model propagate to CR transport and $\gamma$-ray emission.

%% file: Sections/appendixA.tex
\section{CR Spectra for different species}
\label{appendix:spectra}

\begin{figure*}
	\centering
	\includegraphics[width=\textwidth]{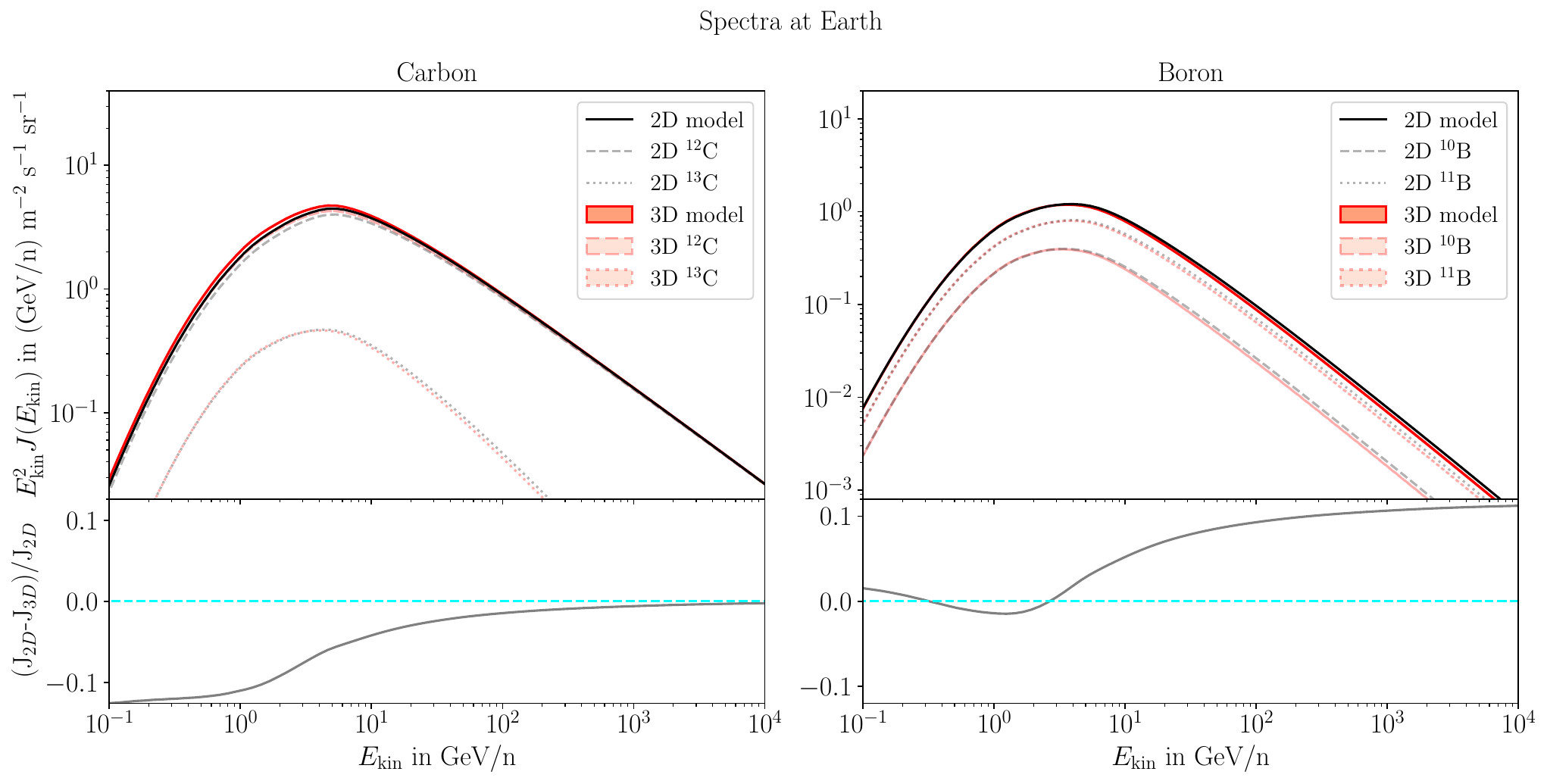}
	\caption{\textit{Top row:} Spectra for the total flux of Carbon (Left) and Boron (Right) in the 2D and 3D models at Earth. Dashed and dotted lines are used for the constituent isotopes as indicated in the legends. Total Carbon and Boron spectra are shown as solid lines. The error band for the 3D model spectra indicates the propagated uncertainty from the different gas samples. Spectra at Earth  are shown with a heliospheric modulation potential of 587.9\,MV.
    \textit{Bottom row:} Corresponding fractional residuals comparing the 3D model to the 2D model.
    }
	\label{BCSpectra1}
\end{figure*}

\begin{figure*}
	\centering
	\includegraphics[width=\textwidth]{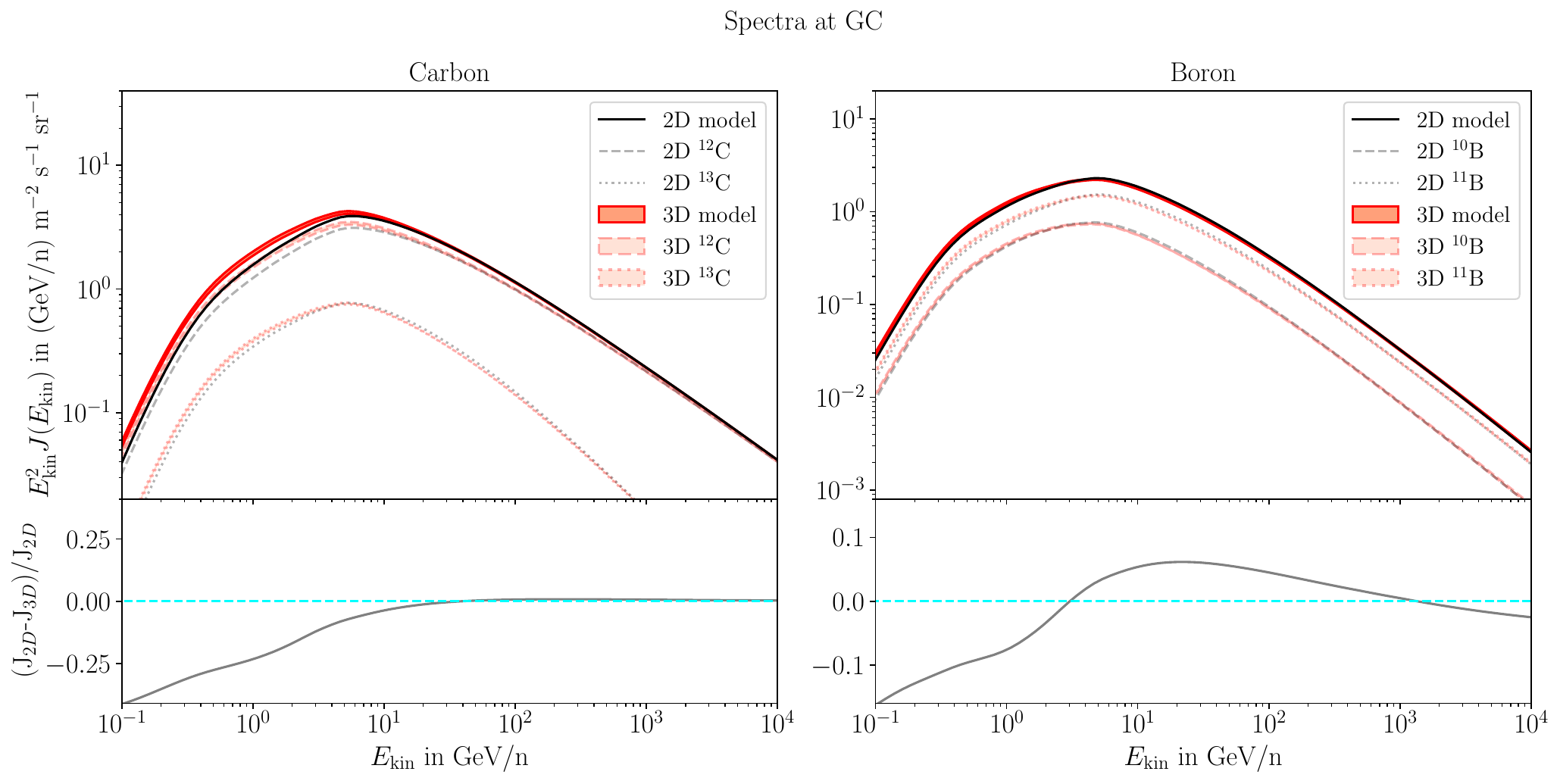}
	\caption{Like Fig \ref{BCSpectra1}, but evaluated at the GC.}
	\label{BCSpectra2}
\end{figure*}

\begin{figure*}
	\centering
	\includegraphics[width=\textwidth]{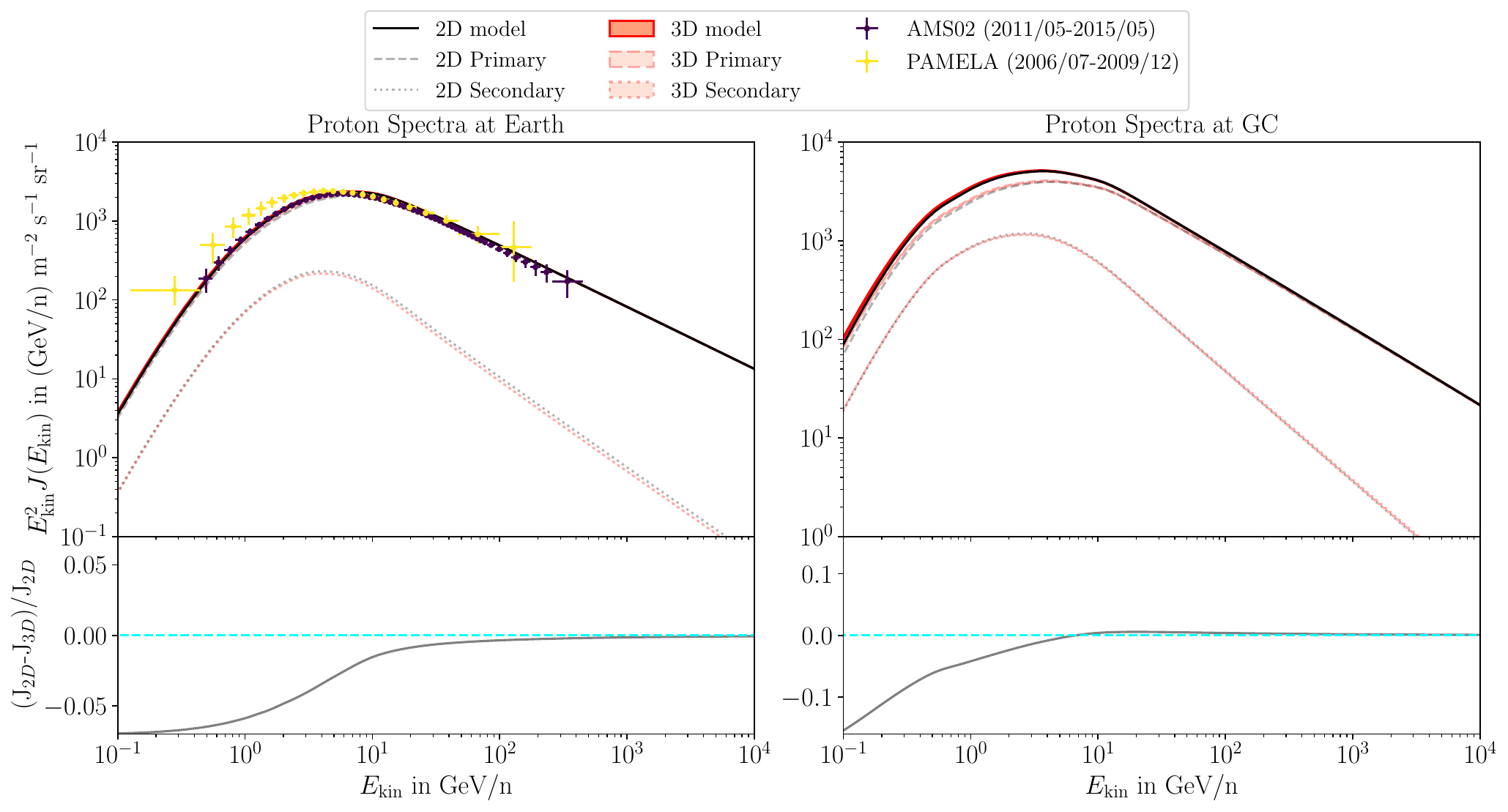}
	\caption{\textit{Left:} Simulated proton spectra at Earth for the 2D model and the 3D model. Primary and secondary components are shown via dashed and dotted lines respectively. Total spectra are shown in solid lines. The error band for the 3D model spectra indicates the propagated uncertainty from the different gas samples. \textit{Right:} Same, but at the GC. Spectra at Earth are compared to AMS-02 \citep{AMS02pData} and PAMELA \citep{PamelapData} data, and shown with a heliospheric modulation of 684.9\,MV. Here, the 3D model parameters are used for both models.
    \textit{Bottom row:} Corresponding fractional residuals comparing the 3D model to the 2D model. Proton spectra residuals are shown at Earth (\textit{left}) and at the GC (\textit{right}).}
    \label{ProtonSpectra}
\end{figure*}

\begin{figure*}
	\centering
	\includegraphics[width=\textwidth]{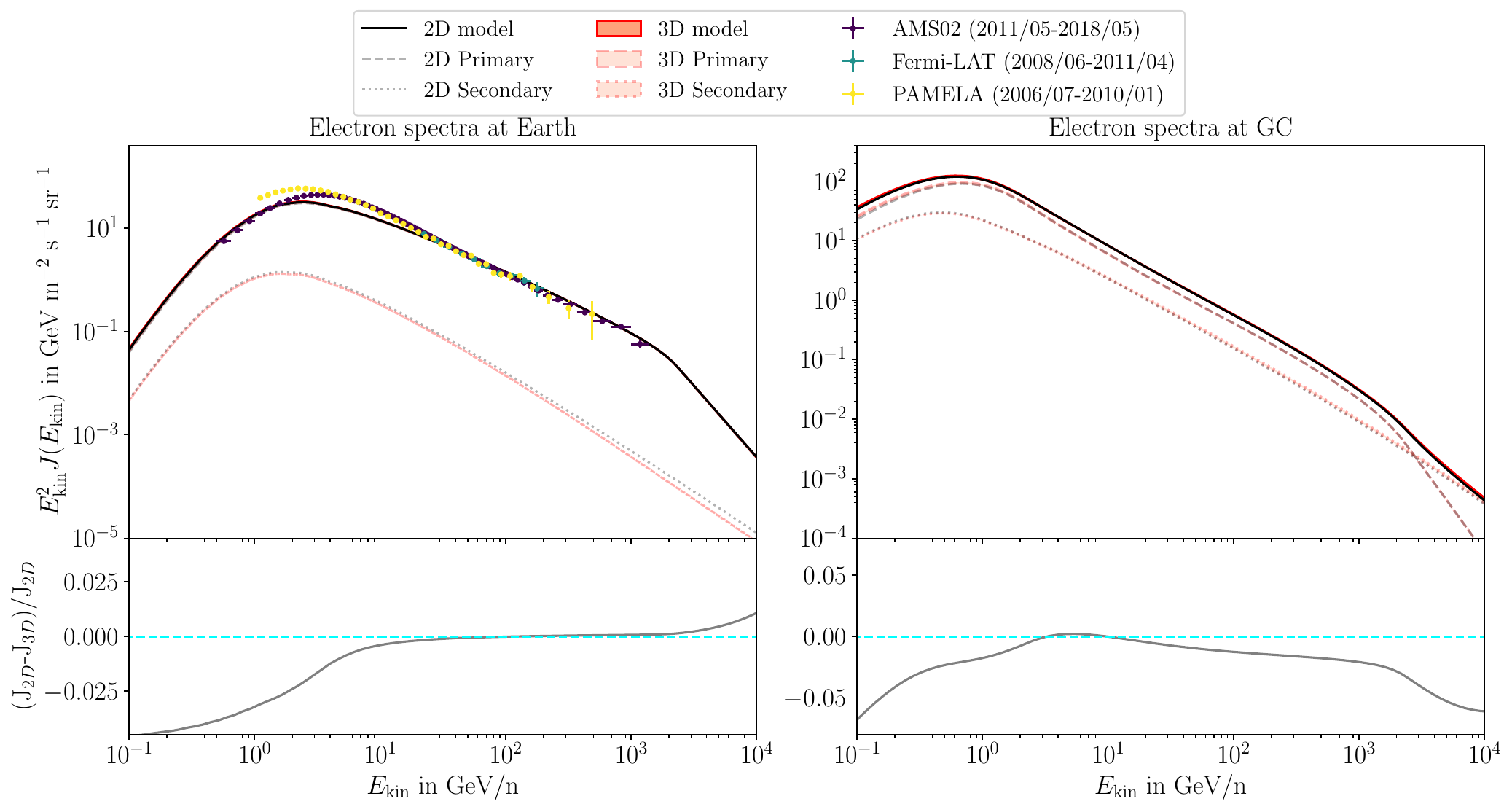}
	\caption{Like Fig \ref{ProtonSpectra}, but for primary and secondary electrons. Spectra at Earth are shown with a heliospheric modulation of 636\,MV and compared to: AMS-02 \citep{AMS02apData}, Fermi-LAT \citep{Fermiedata} and PAMELA \citep{PAMELAeData} data.}
	\label{ElectronSpectra}
\end{figure*}

\begin{figure*}
	\centering
	\includegraphics[width=\textwidth]{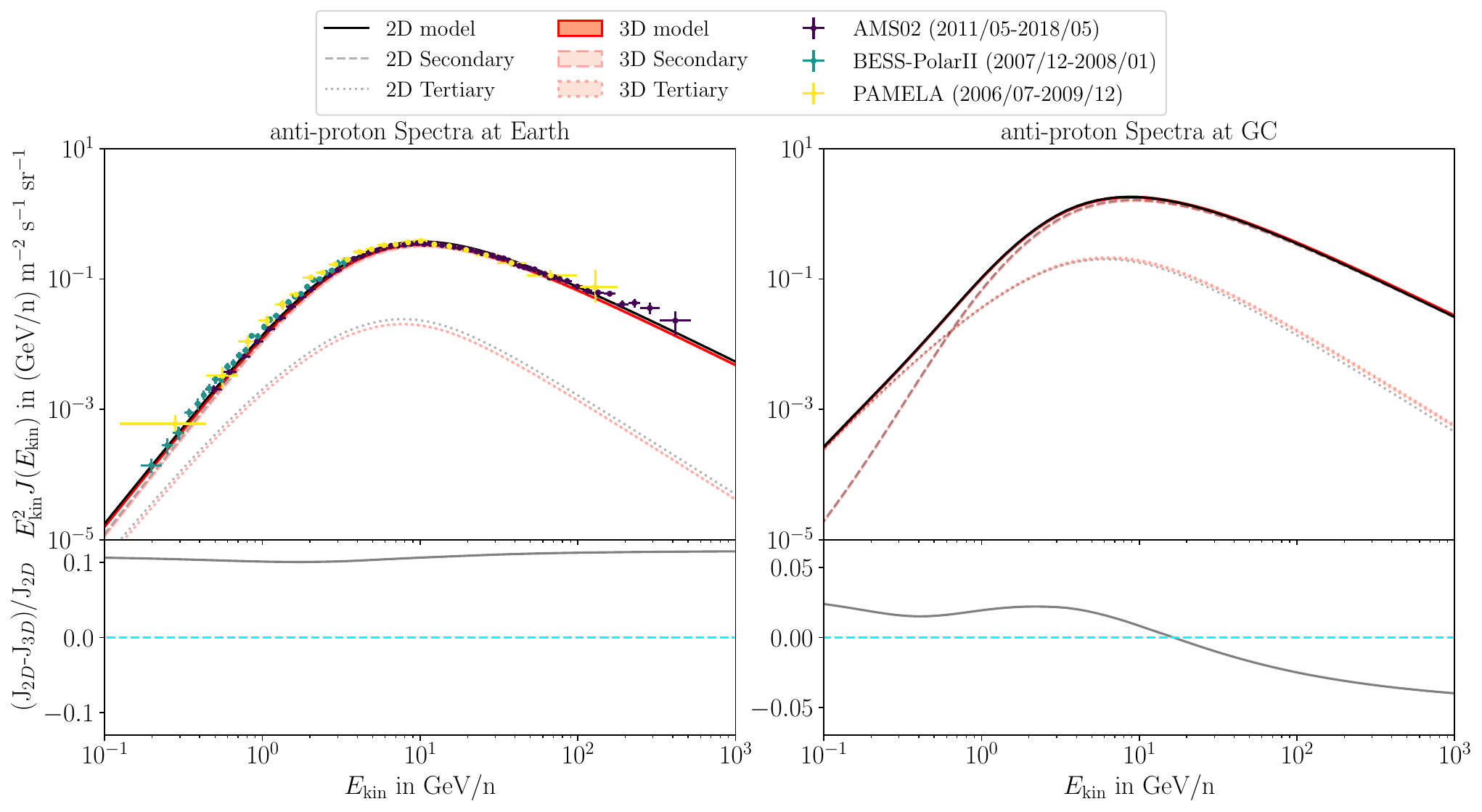}
	\caption{Like Fig \ref{ProtonSpectra}, but for secondary and tertiary antiprotons. Spectra at Earth are shown with a heliospheric modulation of 636.6\,MV and compared to: AMS-02 \citep{AMS02apData}, BESS-PolarII \citep{BESSIIapData}, and PAMELA \citep{PamelapData} data.}
	\label{antiprotonSpectra}
\end{figure*}

Spectra and the corresponding fractional difference between their 2D and 3D model estimations are shown in Figs \ref{BCSpectra1} and \ref{BCSpectra2} (Carbon and Boron), \ref{ProtonSpectra} (Protons), \ref{ElectronSpectra} (Electrons), and \ref{antiprotonSpectra} (antiprotons). 

We find flux differences up to 30\%  at the GC for Carbon at energies below 10\,GeV/n. From Fig \ref{totprot}, we note that at 4\,kpc from the GC there is an under-dense ring in the 2D gas distribution to which CRs can diffuse with equal probability in any direction, whereas in the 3D model the local under-dense regions are located north and south of the GC, constraining the directions in which CRs can diffuse from the GC. This same effect is barely visible for protons, up to 14\% below 10\,GeV/n, and is negligible for electrons, with differences up to 5\% below 10\,GeV/n. However, for electrons, secondaries dominate the spectrum above $\sim2$\,TeV at the GC, due to the softening of the primary injection spectrum, leading to differences increasing with energy above 10\,GeV/n. 
Like for secondary electrons, Boron also exhibits increasing differences with energy, however, in this case the 3D model predicts fluxes up to 10\% lower than the 2D model. 
Although Carbon and Boron are anti-correlated, their local abundances can vary differently with energy due to spallation processes, since their relevant production channels are not identical. For instance, despite ${}^{12}$C being the most important production channel for ${}^{11}$B, ${}^{16}$O dominates the production of Carbon via ${}^{13}$C (followed by ${}^{12}$C) which contributes weakly to the production of B isotopes \citep{prodcrosssections}.
Furthermore, every species incurs spallation and energy losses that are modulated by the local gas distribution, as discussed for protons in Section \ref{subsec:correlations}. Therefore, the anti-correlation between Carbon and Boron observed in our simulations is modulated by such processes and varies with energy.

For protons and electrons, the differences between models are negligible, which comes as a consequence of primaries dominating the total flux, and primary protons and electrons being used for the normalization of the fluxes. We note the discrepancy between the total electron spectrum and AMS-02 data, which can be addressed by adapting the injection spectra of electrons that were used in the simulations. 

Antiprotons probe different spatial scales of the ISM than light elements such as carbon and boron, and are largely correlated to the parameter space of the propagation model \citep{Johannesson_2016}. Being entirely produced through interactions of CR nuclei with Galactic gas, they are heavily influenced by the distribution of the latter. In Fig \ref{antiprotonSpectra} we show the predicted antiproton spectra at Earth and at the GC. As with protons, we note that our predictions are in good agreement with available data. Here we find a $\chi^2$ value of 1.89 above 10\,GeV, despite the fact that antiprotons are not used to optimize the transport parameters in this work. We note that the high gas density at the GC increases the production of tertiary antiprotons, causing them to dominate the total antiproton population at energies below $\sim1$\,GeV.

%% file: Sections/appendixB.tex
\section{$\gamma$-ray emission at different regions}
\label{appendix:gammaemission}

\begin{figure*}
	\centering
	\includegraphics[width=\textwidth]{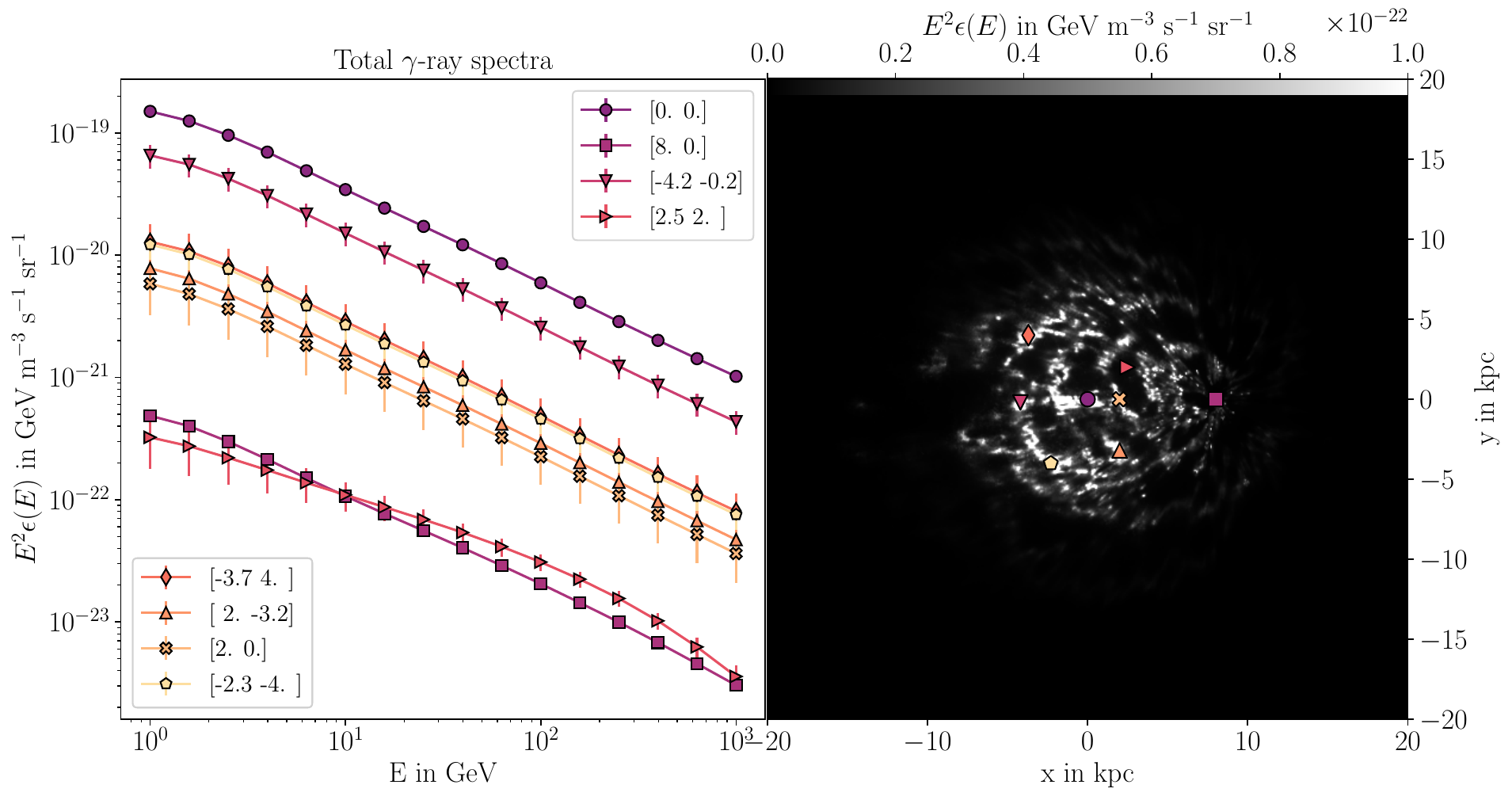}
	\caption{Total $\gamma$-ray spectra simulated at specific locations of varying density in the Galactic plane.}
	\label{GammaSpectraLoc}
\end{figure*}

In Fig \ref{GammaSpectraLoc} we compare $\gamma$-ray spectra from regions of varying gas density, where we show the simulated mean total $\gamma$-ray flux together with the corresponding uncertainties at different locations in the Galactic plane. At locations with locally low gas density, the emission is not directly correlated to the gas density, as is usually the case, because IC emission dominates over the other emission channels. For example, at $(x,y,z)=(2.5,2,0)$ where the gas density is as low as $n_{3D}(2.5,2,0)=0.16$\,cm$^{-3}$ we find a higher total $\gamma$-ray emission than in the vicinity of Earth with gas density $n_{3D}(8,0,0)=0.65$\,cm$^{-3}$. We also find that at the aforementioned low-density region, the total $\gamma$-ray emission exhibits a change in slope above 10\,GeV when compared to the emission at the other locations, as seen before for IC emission.

\begin{figure*}
	\centering
	\includegraphics[width=\textwidth]{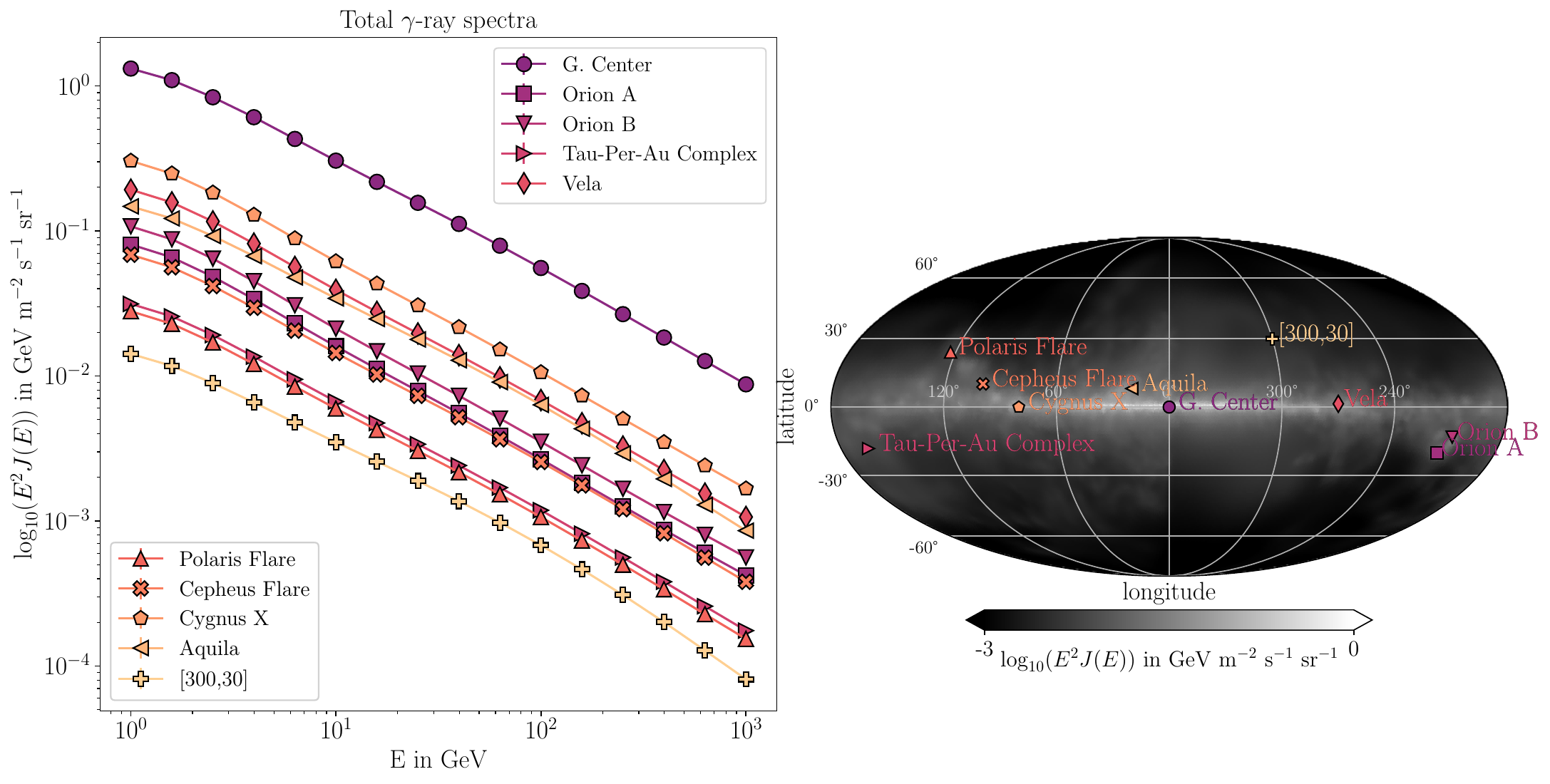}
	\caption{\textit{Left:}LOS-integrated total $\gamma$-ray spectra emitted from specific directions in the sky. \textit{Right:} All-sky map for total $\gamma$-ray emission with the locations of the relevant directions highlighted.}
	\label{GammaSpectraLos}
\end{figure*}

In Fig \ref{GammaSpectraLos} we show $\gamma$-ray spectra from different directions in the sky. We particularly choose some star-forming regions and giant molecular clouds, along with a high-latitude under-dense region. Here, the total $\gamma$ emission is again proportional to the gas density, due to the dominance of $\pi^0$-decay emission.

%% file: Sections/appendixC.tex
\section{$\gamma$-ray emission differences due to gas target}
\label{appendix:Gamma-model-diffs}

The computation of CR transport and $\gamma$-ray emission can be performed with a different gas map for each in our setup. Therefore, our model supports different options for computing the $\gamma$-ray emission, of which we note 3 relevant ones:

\begin{itemize}
    \item CR transport and $\gamma$-ray emission using the 3D gas distribution (3D model)
    \item CR transport and $\gamma$-ray emission using the 2D gas distribution (2D model)
    \item CR transport using the 2D gas distribution and $\gamma$-ray emission using the 3D gas distribution (2D-3D model)
\end{itemize}

\begin{figure}
    \centering
    \includegraphics[width=\linewidth]{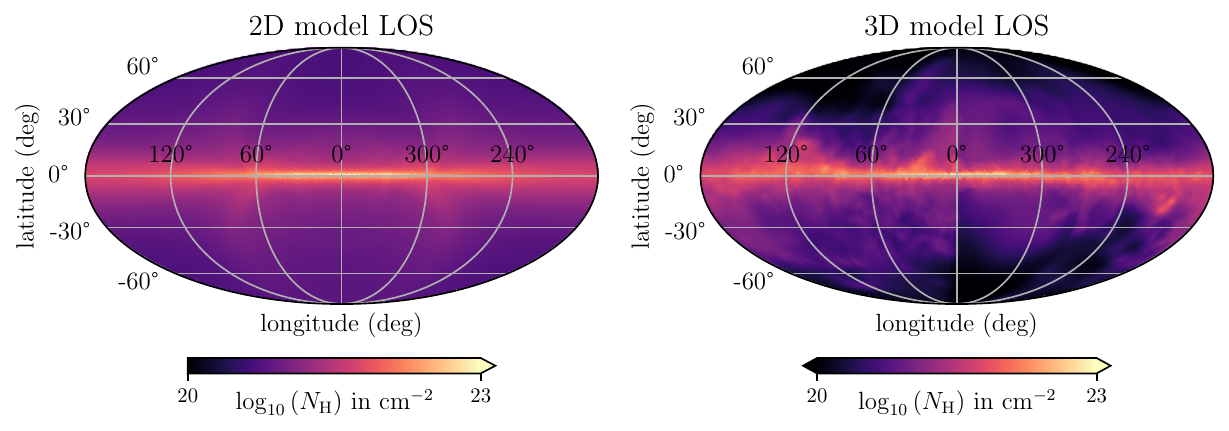}
    \caption{All sky map for the total gas distribution corresponding to the 2D model (left) and 3D model (right).}
    \label{2D3Dtotalgas}
\end{figure}

\begin{figure}
    \centering
    \includegraphics[width=\linewidth]{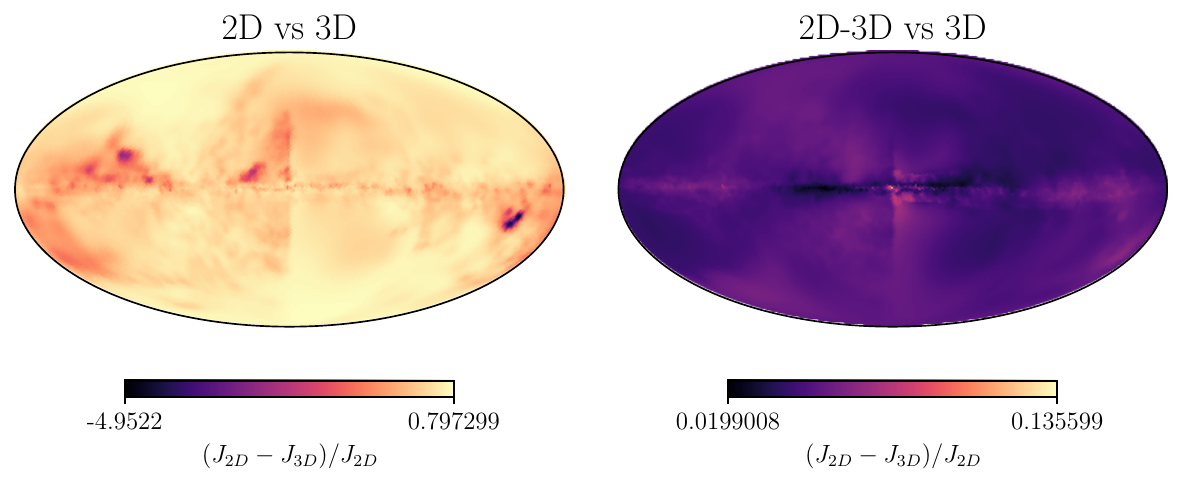}
    \caption{Fractional residuals comparing the total $\gamma$-ray emission at 1\, GeV calculated using the 2D model (left) and 2D-3D model (right), against the one calculated using the 3D model.}
    \label{gammaresiduals}
\end{figure}

In Fig. \ref{2D3Dtotalgas} we show the corresponding sky maps for the total gas distribution in the 2D and 3D gas models. The structures seen extending to higher latitudes in the 2D model come as a consequence of gas clouds localized entirely behind the GC in the 3D gas maps contributing to densities at higher latitudes when averaging to produce the 2D model. We show the fractional differences for both the 2D and 2D-3D models versus the 3D model in Fig. \ref{gammaresiduals}. As expected, the 2D case deviates considerably from the 3D model, especially at dense cloud regions, at which the gamma-ray flux would be below the 3D predictions. Furthermore, the 2D-3D case varies moderately, to the same degree as the CR spectra, although it exhibits a larger deviation in the vicinity of the GC.

